\documentclass[12pt,preprint]{aastex}

\slugcomment{}

\shorttitle{Nearby Galaxies in 2MASS}
\shortauthors{Devereux et~al.}

\begin{document}

\newcommand{\etal}{{et~al.}~}


\title{Nearby Galaxies in the 2${\micron}$ All Sky Survey\\
I. $K$-band Luminosity Functions}


\author{Nick Devereux\altaffilmark{1} }
\affil{Department of Physics, Embry-Riddle Aeronautical University,
    Prescott, AZ 86301}
\email{devereux@erau.edu}

\author{S.P. Willner \& M.L.N. Ashby}
\affil{Harvard-Smithsonian Center for Astrophysics}
\email{swillner@cfa.harvard.edu, mashby@cfa.harvard.edu}

\author{C.N.A. Willmer}
\affil{University of Arizona}
\email{cnaw@as.arizona.edu}

\and

\author{Paul Hriljac }
\affil{Department of Mathematics and Computer Science, Embry-Riddle Aeronautical University,
    Prescott, AZ 86301}
\email{hriljap@erau.edu}    


\altaffiltext{1}{Fulbright Scholar, Center for Astronomy, National University of Ireland, Galway}


\begin{abstract}

Differential $K_{s}$-band luminosity functions (LFs) are presented for a complete sample of 1613 nearby bright galaxies segregated by visible morphology. 
The LF for late-type spirals follows a power law that rises towards low luminosities whereas the LFs for ellipticals, lenticulars and bulge-dominated spirals are peaked
and decline toward both higher and lower luminosities.
Each morphological type (E, S0, S0/a--Sab, Sb--Sbc, Sc--Scd) contributes approximately equally to the overall $K_{s}$-band luminosity density of galaxies in the local universe.
Type averaged bulge/disk ratios are used to subtract the disk component leading to the prediction that the $K_{s}$-band LF for bulges is bimodal with ellipticals dominating
the high luminosity peak, comprising 60\% of the bulge luminosity density in the local universe with the remaining 40\%  contributed by lenticulars and the bulges of spirals. Overall, bulges contribute 30\% of the galaxy luminosity density at $K_{s}$ in the local universe with spiral disks making up the remainder.
If bulge luminosities indicate central black hole masses, then our results predict that the black hole mass
function is also bimodal.

 \end{abstract}


\keywords{galaxies: luminosity function, mass function --- galaxies: formation --- galaxies: elliptical and lenticular, cD --- galaxies: spiral --- galaxies: bulges --- infrared: galaxies}



\section{Introduction}

Luminosity functions (LFs) are one of the key statistical instruments widely used 
to better our understanding of galaxies, because with 
just a few parameters, they concisely describe entire populations of 
objects whose intrinsic
properties (e.g., mass, luminosity, etc.) vary over orders of magnitude.  
In the Schechter (1976) formulation, those parameters are 
$M^*$, the absolute magnitude corresponding to the knee of the LF, $\phi^*$, the mean galaxy space density,
and $\alpha$, the faint-end slope of the LF.
The basic technique is to measure and then compare these parameters 
for different populations and thereby assess,  for example, how galaxies evolve 
with redshift, or as a function of environment, or indeed any other 
variable that can be controlled for. This tool has been used, as of late, to analyse
several large samples of galaxies reaching out to redshifts $z\sim$1
and beyond \cite[e.g.,][]{Nor02, Bla03, Wol03, Bel03, Ilb05, Wil06, Bro07, Fab07, Wak06, Red08}.

Galaxy morphology is a taxonomy devised by 
\citet{Hub36} and refined by \citet{deV59} and \citet{deV91} that is based on the relative 
prominence of the stellar bulge and the degree of resolution of the spiral arms.  
The physical significance of galaxy morphology is that it reflects the 
galaxy merger history \citep[e.g.,][]{Balland1998}. It is conventional wisdom that 
the galaxy LF does depend on morphology in the $B$-band. \citet{San85} found that LFs depend on morphological type for galaxies in the Virgo cluster and subsequent works showed that this dependence extends to
field galaxies \citep{Bin88,Lov92}. However, \citet{Efs88} and \citet{Mar94,Mar98} offer a contrary view. 
More recently, LFs based on morphological proxies\footnotemark,  \footnotetext{Morphological proxies such as color are adopted until the 
difficult and time consuming task of assigning morphologies to the 
ever-growing number of cataloged galaxies can be accomplished e.g.,The Galaxy Zoo Project, http://www.galaxyzoo.org/} such as color,
also show
a dependence 
\citep[e.g.,][]{Bla01,Bel04, Bla06, Wil06,
Fab07}, at least at visible wavelengths.  These studies and others indicate
that the present-day LFs are the result of a complex evolutionary
history.

By virtue of its uniformity, reliability, and full-sky coverage, 
the Two Micron All Sky Survey \citep[2MASS,][]{Skr06} has recently made it
possible to extend LF studies into the infrared regime.  
2MASS has been exploited in recent years to produce near-infrared luminosity 
functions for galaxies with ever greater 
precision, facilitated by redshifts generated from the SDSS 
 \citep{Yor00} and the 2 and 6 degree Field Galaxy Redshift Surveys 
\citep{Col01, Koc01, Bel03, Eke05, Jon06}.  The  2.16${\micron}$
$K_{s}$-band (hereafter $K$-band) LF 
is of particular interest because of its relevance to
understanding galaxy evolution in the context of Lambda Cold Dark Matter 
(${\Lambda}$CDM) cosmology: at $z=0$, the $K$-band light traces 
the stellar mass accumulated in galaxies at a wavelength where interstellar 
extinction is minimal \citep{Dev87, Bad01, Bel03}, avoiding strong dependence on metallicity
and stellar population age.  For this reason, $K$-band
LFs complement those obtained at visible wavelengths. 

Both \citet{Koc01} and \citet{Bel03} investigated the impact of morphology on the
$K$-band LF.  \citet{Bel03} used an SDSS concentration parameter 
($c_r$) as a proxy for type. \citet{Koc01} carried out a painstaking
typing procedure in such a way as to permit a quantitative understanding 
of the uncertainties involved.  Both studies divided the galaxy samples 
into two broad categories; early and late.  They arrived at consistent 
total $K$-band LFs  \citep[][Figure~9]{Bel03}, 
and found that 
the early- and late-type LFs were well-described by 
Schechter functions that differed slightly in $M^*$ and $\phi^*$, but had essentially identical shapes. 
 This is somewhat surprising
given the distinctions revealed by LFs measured in the visible.
Clearly more work remains to be done on the morphological type dependence 
of LFs in general, and for the $K$-band LF 
in particular. Fortunately, the time-consuming task of assigning visible morphologies has 
now been completed for the vast majority of nearby galaxies.  The principal aim 
of this paper is therefore to use these nearby galaxies to define new benchmark 
$K$-band LFs for galaxies.  As such, our study offers an improvement on  \citet{Koc01} and \citet{Bel03} by using the most recent distances 
and {\em finely divided} 
morphological types.  

The paper is organized as follows.  Section 2 describes how the 
nearby galaxy sample is selected.  A non-parametric (Choloniewski) method is used to 
generate the LFs as described in section 3.  The results, presented 
in section 4, include the morphological type dependence 
of the $K$-band LFs as well as the contribution of each 
to the luminosity density in the local universe. Recently published data have also provided
the opportunity to compute the bulge luminosity functions and hence the contribution of bulges
and disks to the $K$-band luminosity density in the local universe. These results are
also presented in section 4. A discussion, in section 5, explains how the results may be used 
to constrain models of galaxy evolution and includes a prediction that the bulge LF
and by association, the black hole mass 
function, is bimodal and depends on Hubble type. Conclusions follow in Section 6.

\section{ Sample Selection}

The goal of this study is to quantify the dependence of the $K$-band LFs on the visual appearance of galaxies. This dictates that the sample be composed of nearby galaxies which have 
the most reliable morphological assignments.  Nearby galaxies were identified using HYPERLEDA; a web-based interface 
(http://leda.univ-lyon1.fr) that provides access to the Principal Galaxy 
Catalog \citep[hereafter PGC,][]{Pat03}.  The PGC is a homogeneous database 
of galaxy parameters in the sense that an attempt has been made to place 
independent measures on a standard system.  Parameters employed in this study 
include coordinates accurate to ${\sim}2$\arcsec, visible morphological {\tt T} types 
and recession velocities for all known galaxies brighter than $B=18$\,mag.  
The PGC is a dynamic resource that is constantly being updated, but as of 
2008 January there were 7406 nearby galaxies with V$_{\mathrm gsr}$ ${\le}$ 3000\,km\,s$^{-1}$,
apparent total blue magnitudes m$_{B}$ ${\le}$~18\,mag and Galactic 
latitudes ${|b|}$  ${>}$ 10 degrees.  The latitude constraint was imposed 
to avoid the inevitable incompleteness due to obscuration within our galaxy.
However, with the exception of the Galactic plane, the PGC galaxies from which the sample is selected are distributed over the entire sky thereby minimizing the effect of cosmic variance due to large scale structures present in the sample volume.

The 2MASS counterparts of the PGC galaxies were identified on the basis of positional coincidence. 
For an association to be made, the PGC J2000 coordinate had to fall within 10${\arcsec}$ of the 2MASS J2000 coordinate listed in the Extended Source Catalog\footnotemark \citep[XSC,][]{Jar00}. 
\footnotetext{Final checking revealed 30 galaxies with PGC and XSC coordinate differences greater than 10${\arcsec}$. The discrepancy is due to different centroids in visible and infrared light which is a problem particularly for nearby galaxies of large angular size.
These galaxies are included in Table~1.}This comparison yielded 5034 detections  at 2.2\,${\micron}$ ($K$-band) corresponding to a 68\% detection rate. Figure 1 shows the distribution of apparent $K$-band isophotal magnitudes for nearby galaxies (V$_{\mathrm gsr}$ ${\le}$ 3000\,km\,s$^{-1}$) in the 2MASS XSC. The distribution turns over at $K$ = 10\,mag indicating that the XSC is incomplete for galaxies fainter than that. On this basis,  a volume-limited sample was defined for further study, hereafter the K10/3000 sample, comprising 1613 galaxies with ${K}$  ${\le}$ 10\,mag, V$_{\mathrm gsr}$ ${\le}$ 3000\,km\,s$^{-1}$,  and ${|b|}$  ${>}$ 10 degrees (Table~1). The choice to use V$_{\mathrm gsr}$ allows us to define a spherical volume centered on the Milky Way that simplifies the LF calculation.

Every attempt has been made to identify all known galaxies with $K$ ${\le}$ 10\,mag, V$_{\mathrm gsr}$ ${\le}$ 3000\,km\,s$^{-1}$,  and ${|b|}$  ${>}$ 10 degrees. 
Nevertheless, sample incompleteness can arise in two ways: 1) galaxies that are in
the 2MASS XSC but have no radial velocities in the PGC or are missing
from the PGC altogether, and 2) PGC galaxies missing from the XSC.  A
firm upper limit on the former can be set by noting that there are 388
objects with $K{\le} 10$ and ${|b|}$  ${>}$ 10 degrees in the 2MASS XSC for which no association with
a PGC galaxy can be found.  Inspection of 2MASS images reveals that
most are Galactic star clusters (or indeed in some cases small parts of
star clusters treated by the automated XSC extraction procedure as
distinct objects), but some are galaxies.  Redshifts have been
determined by J. Huchra, L. Macri, T. Jarrett, J. Mader, A. Crook,
R. Cutri, T. George, N. Martimbeau,  S. Schneider, \& M. Skrutskie
(ApJ Supp., 2009, in preparation) for all 166 of the galaxies.  
All but three of which have
radial velocities $>$3000\,km\,s$^{-1}$.  The three with
$V<3000$ km\,s$^{-1}$ should in principle be included in the K10/3000
sample but are not because they have no entries in the PGC; this gives an
estimated completeness for Table~1 of 99.8\%.  With regard to the second
source of incompleteness, there are two well known PGC galaxies of large angular size
missing from the 2MASS XSC: 
the Large and Small Magellanic Clouds. These 
and some other local group dwarf galaxies are not included in the XSC \citep{Jar03} and hence are not included in the K10/3000 sample. However,
it is unlikely that there are any PGC spiral and elliptical galaxies
missing from the K10/3000 sample because the $K=10$ sample limit corresponds to $B=13.5$ for
a normal $B-K=3.5$ galaxy color, which is 4.5 magnitudes brighter than the
PGC catalog limit.  Thus, only galaxies with very blue colors and/or very low
surface brightness would be missing from our sample \citep{And02} and both
selection effects will tend to further bias our sample against late-type (Sd and later) spirals and
dwarf irregular (Im) galaxies, as explained in more detail in the next section.

\subsection{Morphological Types in the K10/3000 Sample}

The observed distribution of $K$ magnitudes for galaxies in the XSC with V$_{\mathrm gsr}$ ${\le}$ 3000\,km\,s$^{-1}$ is compared with the distribution of blue magnitudes for the entire inventory of 7406 PGC galaxies contained within the same volume (Figure 1). The figure shows that 2MASS detected essentially the same number of nearby galaxies with $K{\le}$ 
10\,mag as would be found in a $B$-band selected sample with $B{\le}$ 13.5\,mag.  

Figure 2 illustrates how the $B-K$ color distribution for a $K$-band limited sample differs from a $B$-band selected one. Although a $B$-band selected sample with $B{\le}$ 13.5\,mag
contains about as many galaxies as a $K$-band selected sample with ${K}$  ${\le}$ 10\,mag, the latter sample excludes blue galaxies contained between the two magnitude limits, figuratively speaking, between 6 and 8 o'clock on the plot. However, the deficiency is almost exactly compensated for by the inclusion of additional red galaxies between 12 and 2 o'clock. Thus, a $K$-band magnitude limited sample will contain more red and fewer blue galaxies compared to a $B$-band selected sample of similar size.

Since a $K$-band selected sample contains fewer blue galaxies compared to a $B$-band selected 
sample of comparable size, one can therefore anticipate that morphological types in the K10/3000 sample will contain fewer star forming late-type spirals and dwarf irregular galaxies, compared to a $B$-band selected sample of similar size. Figure 3 supports this expectation. The distribution of morphological types in the K10/3000 sample spans the entire range from ellipticals to dwarf irregulars.
However, compared to the sample of PGC galaxies with ${B}$ ${\le}$ 13.5\,mag, 
the K10/3000 sample appears to be missing more than 50\% of galaxies
with morphological types T ${>}$  6, corresponding to Sd and later.  Consequently, these 
types of galaxies will not be considered further as there are too few to reliably define a LF. On the other hand, the $K$-band is sensitive to the red luminous mass
component in all galaxy types with the result that the K10/3000 sample contains 80\% of the Sc--Scd types and all the earlier types, plus more, than would be found in a $B$-band selected sample of comparable size.

A completeness test that makes no
assumptions concerning the galaxy distribution and is unaffected by the presence of
large scale structure was proposed by \cite{Rau01}. 
The test assumes that the LF of the population does
not depend on the three-dimensional redshift-spatial distribution and that
the apparent magnitude limit can be described by a sharp cutoff. In
practice the test examines the distribution of the random variable
\begin{equation}
\zeta = \frac {\int_{-\infty}^M \phi(M) dM} {\int_{-\infty}^{M_{lim}(z)}\phi(M) dM}
\end{equation}
which measures the ratio between the integrated LF
up to the absolute magnitude of a given galaxy  and the total range in
magnitudes accessible at the distance of the galaxy.
Completeness is measured by the variable
\begin{equation}
T_c = \frac {\sum_{i=1}^{N_{gal}} (\zeta_i - 0.5)} {(\sum_{i=1}^{N_{gal}} V_i)^{1/2}}
\end{equation}
where the $V_i$ represent the variance of the $\zeta_i$ estimators. The
$T_c$ variable has an expectation value of zero and unit variance.
In practice
the test is evaluated for sub-samples selected at progressively fainter apparent magnitude limits. 
The $T_c$ statistic will fluctuate around the value of zero, but
becomes systematically negative once the sample becomes incomplete. From
the behavior of this statistic the limiting magnitude can be inferred.
Figure 4 illustrates the result of applying this test to the
full K10/3000 sample and to several sub-samples
segregated by morphological type. Both the number counts and the Rauzy tests 
suggest that the current sample
is not affected by incompleteness for the morphological types we are
considering, namely ellipticals, lenticulars and spirals up to and including types Scd.

\subsection{Galaxy Distances}

Because the sample galaxies are nearby, their peculiar velocities can be a significant fraction of their Hubble flow velocities, particularly for those with  V$_{\mathrm gsr}$ ${\le}$ 1000\,km\,s$^{-1}$. Major
perturbers include the Virgo
cluster, which lies inside the sample volume, and the Great Attractor \citep{Lyn88}
which lies outside the volume but still perturbs the Hubble flow within. Thus, one can not simply deduce a distance from the recession velocity and a Hubble constant.  \cite{Tul08} have recently quantified the peculiar velocities for nearby galaxies utilizing redshift-independent distances based on the
following methods; the Tully-Fisher relation \citep{Tul77}, Cepheids \citep{Fre01}, the luminosity of stars at the tip of the red giant branch \citep{Kar04,Kar06}, and surface brightness fluctuations \citep{Ton01}. These methods have yielded {\it quality} distances, with distance modulus uncertainties ${<}$ 0.1 mag, for 591 nearby 
(V$_{\mathrm gsr}$ ${\le}$ 3000\,km\,s$^{-1}$) galaxies of which 302 are included in the K10/3000 sample. Tully determined distances for other galaxies based on associations with groups which contain one or more members with quality distances together with the Numerical Action Models of \cite{Sha95}. Collectively, datasets provided by Tully (2007, private communication) yielded  distances for 1575  or 98\% of the 1613 objects in the K10/3000 sample and they are listed in Table~1\footnotemark.

\footnotetext{See also The Extragalactic Distance Database (EDD) at http://edd.ifa.hawaii.edu/}

\subsection{2MASS magnitudes}

In the following analysis,
isophotal magnitudes (measured within the 20\,mag\,arcsec$^{-2}$ elliptical isophote; the parameter: {{\tt  k${_{-}}$m${_{-}}$k20fe} in the 2MASS XSC) are adopted to characterize the dependence of $K$-band
LFs on galaxy morphology (section 4.2). The main reason for this choice is that the extrapolated total magnitudes published in the 2MASS XSC are unreliable for elliptical galaxies due to a restriction on the choice of S\'ersic index that caused their total flux to be underestimated \citep{Lau07}, by an amount we determine
to be ${\sim} 0.3$ mag\footnotemark \footnotetext{The average difference between the isophotal and total magnitudes quoted in the XSC is ${\sim}$ 0.1 mag for K10/3000 elliptical galaxies. However, based on an analysis described in section 4.4, we find that the total magnitude is likely to be 0.3 mag brighter, on average, than the total magnitude quoted in the XSC for K10/3000 elliptical galaxies.}. Additionally, extrapolated total magnitudes in the 2MASS XSC can lead to unphysical colors  \citep{Kar02} suggesting that the extrapolations are unreliable for some spiral galaxies as well. With the exception
of the ellipticals, the mean difference between the isophotal and total magnitudes 
cited in the 2MASS XSC is small: about 0.14 mag for the bright, ${K}$  
${\le}$ 10\,mag, galaxies in our sample and is independent of galaxy type for morphologies 
spanning S0 to Scd. Thus, the isophotal magnitudes
are adopted as published with no corrections with the understanding that they slightly underestimate
the total magnitudes for spiral and lenticular galaxies but significantly underestimate the total magnitudes for
ellipticals: a detail that is addressed in sections 4.3 and 4.4.

\section{Luminosity Function Determination}

The LF calculation uses the maximum likelihood method
of \cite{Cho86} which assumes that the spatial distribution of galaxies and their luminosities are uncorrelated. 
The merit of this common Poisson assumption is discussed further in the Appendix. 
By counting galaxies
in a plane defined by distance and luminosity it is possible to obtain,
simultaneously, the density distribution as a function of distance and
the properly normalised LF, unaffected by density
variations within the sample \citep{Cho86, Tak00}. The Choloniewski method is non-parametric, i.e., it makes no assumptions of a functional form for the
LF and it yields the overall normalization (galaxy space density) directly. A disadvantage is that it requires binning the data. These characteristics
distinguish it from the maximum likelihood methods of \cite{San79} and \cite{Efs88}.
The interested reader is referred to \cite{Wil97} and \cite{Tak00} for a more detailed inter-comparison of the methods used to calculate galaxy LFs. Generally speaking, the results obtained using the Choloniewski method always agree, within the estimated statistical
uncertainties, with the other methods described by \cite{Bin88, Wil97} and \cite{Tak00}. The same is true for the K10/3000 sample considered here. 

The Choloniewski method is implemented by plotting the two independent quantities; distance modulus and absolute magnitude, as illustrated in Figure 5.  Galaxies are then binned and summed, vertically and horizontally. A small penalty is incurred as a result of the  binning procedure because galaxies contained in partial bins, that are bisected by the apparent magnitude limit of the survey, must be excluded. Additionally, a few galaxies are excluded by the upper and lower bounds of the absolute magnitude and distance modulus limits. Table~2 provides a summary of the binning parameters used to generate the various LFs presented in this paper.
The summations are used to iteratively solve the simultaneous equations 18, 19, and 20 (or 21) cited by \cite{Cho86}. These non-linear equations converge surprisingly quickly to yield ${\phi}$(M),  the differential LF, where M is absolute magnitude, ${\rho}$(${\mu}$), the number density of galaxies as a function of distance modulus, ${\mu}$, and n, the average number density of galaxies in the sample. The procedure was applied to the full K10/3000 sample and also to subsets of galaxies sorted by morphological type.
A detailed explanation of how to correctly populate the covariance matrix, from which the statistical uncertainties are derived, is provided in the Appendix.

In order to test the LF calculation, it was applied to samples of size comparable to the K10/3000 sample drawn from the Millenium Simulation database \citep{Springel2005} and in particular the semi-analytic galaxy catalog within it \citep{DeLucia2007,Croton2006}. For 16 independent samples drawn according to the 
$\mu$ and $m_K$ selection of the K10/3000 sample, the derived LFs agree well with the true one derived by counting all simulated galaxies within the volume. 
The dispersion among the 16 realizations is, however, about 50\% larger than the calculated LF uncertainties, even after normalizing the
16 samples to eliminate cosmic variance. The excess dispersion is also about 50\% larger than the Poisson uncertainties.  Thus, to the extent that the Millenium Simulation is representative of the distribution of galaxies in the local universe, neither the Poisson estimate nor the
Choloniewski estimate is a good representation of the actual uncertainties as measured by the standard deviation of the simulated LFs. The most likely reason for the discrepancy is that real galaxies are clustered. Clustering will increase the uncertainties because the statistically-independent unit consists of multiple galaxies (on average 2.2 of them if the uncertainties are increased by a factor of 1.5).  The effect of clustering is discussed in more detail in the Appendix.  For the Choloniewski method in particular, the discrepancy between the calculated uncertainties and their Poisson values is largest for elliptical, lenticular, and early-type spirals (Table~3) which tend to be the most clustered.  The Choloniewski uncertainty estimates thus seem particularly sensitive to clustering, but future improvements may be possible by implementing the method with a generalized Poisson distribution instead. Overall, the simulation results suggest that the LFs are reliable but the calculated uncertainties are underestimated by about a factor of 1.5 and using the Poisson method to estimate uncertainties would not improve the results. 

\section{Results}

\subsection{Density Function for the K10/3000 Sample}

Figure ~6 illustrates the solid-angle averaged radial number density 
of galaxies at each distance in the sample volume. Since the solid angle average is over very nearly the whole sky (3.3$\pi$ steradians) 
individual structures can account for only a small fraction of any density peak.
For example, the Virgo cluster, which is the largest structure in this volume,  constitutes only ${\sim}$ 14\% of the total galaxy density enhancement seen at ${\mu}$ + 5log$_{10}$${\it h}$ ${\sim}$ 30.5\,mag. 
That peak is likely due to the combination of the Local Supercluster (of which Virgo is part) and
the Southern Supercluster (a Southern hemisphere structure that includes the groups of Dorado and the Eridanus and Fornax clusters). Consequently, one is cautioned against identifying density enhancements with individual known galaxy clusters or groups. 
Clear evidence for a morphology-density relation \citep{Dre80} is therefore absent from the data. 
Late-type spirals (Sc--Scd) have the highest average number density in the sample volume and ellipticals the lowest.

The average density of galaxies appears to trend downwards at ${\sim}$ 30\,Mpc (${\mu}$ + 5log$_{10}$${\it h}$ = 32.38\,mag, $h = 1$), the maximum distance of the sample. The trend is not due to incompleteness for three reasons. First, the Rauzy test results described in Section 2.1. Second, V/V$_{max}$ = 0.7 in the outer shell (31.88 ${\le}$ ${\mu}$ + 5log$_{10}$${\it h}$  ${\le}$ 32.38\,mag) indicating that the sample is complete up to the distance limit of the survey. Third,  \cite{Cho86} reported the same trend using an independent sample based on the CfA redshift survey. That sample revealed that the density increases again farther out at ${\sim}$ 80\,Mpc ($h=1$) (${\mu}$ + 5log$_{10}$${\it h}$ = 34.51\,mag). Thus, the downward trend at the periphery of the K10/3000 sample is judged to be real and not symptomatic of a selection effect. 

The decrease in galaxy density at 30\,Mpc ($h=1$) is large but within the plausible range of cosmic variance.  An analytic estimate of cosmic variance \citep{dh} depends on the volume sampled and the
galaxy two-point correlation function.  The latter is not known directly for the $K$-selected sample, but we assume it's the same as
for visually-selected galaxy samples.  A correlation function based
on the fluctuation power spectrum from WMAP--1 \citep{Spergel2003} was
extrapolated to the present via the method of \citet{Seljak1996} then
transformed via a spherical Bessel function and smoothed with a 1~Mpc
radius (D.~Eisenstein, private communication, 2006).  The volume
integral \citep[e.g.,][equation~1]{Newman2002} was then evaluated via a
Monte Carlo approach.  The resulting cosmic variance uncertainty is
$\sim$12\% for the full sample and $\sim$18\% for a sample half
a magnitude less deep (Figure~6).  The true correlation function is
likely to be larger for $K$-selected galaxies, which are
predominantly early-type, than for visually-selected ones, more of
which are late-type (Figure~3).  This suggests the preceding
estimates may be too small.  An alternate estimate for cosmic
variance makes further use of the Millennium Simulation
\citep{Springel2005} and the semi-analytic galaxy
catalog \citep{DeLucia2007,Croton2006}.  For this calculation,
simulated galaxies were simply counted in 64 independent spheres of
30 and 23.8~Mpc ($h=1$) radii, and standard deviations were 26\% and
33\% respectively of the mean galaxy density.  These values would
make the dip in the final bin of Figure~6 (or the excess in preceding
bins) only about 2$\sigma$.  The agreement of our overall luminosity
function with that of \citet{Jon06} (section 4.2) suggests that cosmic variance is
not playing a large role in our results.  Regardless of its
magnitude, cosmic variance represents a single uncertainty for the
entire LF, not an independent uncertainty in each
luminosity bin, under the assumption that the LF is independent of location.

\subsection{Parametric Fits to the $K$-band Luminosity Functions by Hubble Type}

Figure 7 presents $K$-band LFs with Schechter function fits \citep{Sch76} for
all galaxies and subsets segregated by morphological type.  The LFs
are defined in Table 3 and the corresponding Schechter function fit parameters are listed in Table~4.
For the total $K$-band LF, values for the parameters $M_{*}$$-$5log$_{10}{\it h}$, 
$\phi_{*}$/${\it h}$$^{3}$  and $\alpha$
agree within 2${\sigma}$ of previous determinations 
\citep{Jon06, Eke05, Bel03, Koc01, Col01}. 
We find a ${\sim}$ 40\% higher space density of galaxies in the range 
$-$23 $<$ M$_{K}$$-$5log$_{10}{\it h}~{<}$ $-$21\,mag than \cite{Jon06} which represents a 
difference of about 2${\sigma}$. Thus, unlike \cite{Jon06}, we do not find a residual with respect to the best fitting Schechter function over that magnitude interval.

Figure~7 shows that the $K$-band LF for ellipticals is represented by a Schechter 
function that declines toward both high and low luminosities, akin to the {\it red} elliptical sequence of \cite{Dri07}. Number counts decrease for low-luminosity elliptical galaxies until the low-luminosity upturn at M$_{K}$$-$5log$_{10}{\it h}>$ $-$21\,mag which represents the onset of the dwarf elliptical sequence at $M_{B} >$ $-$18 \citep{San85}, 
akin to the {\it blue} elliptical sequence of \cite{Dri07}. 

The LFs for lenticular galaxies (S0) and 
bulge-dominated early-type spirals (S0/a - Sbc) are very similar and therefore have been combined.
Like the ellipticals, the combined luminosity 
function for lenticulars and bulge-dominated spirals is represented by a Schechter 
function that declines toward both high and low luminosities. 
In contrast, the LF for late-type spirals (disk-dominated; types Sc--Scd) is completely different than found for the other galaxy types.  Although this LF is also is well-represented by a Schechter function, 
the function is 
{\em essentially} a power law over the range of luminosities for which the LF can be defined, with a slope ${\alpha} = $$-$1.4 that predicts an increasing space 
density of low-luminosity late-type spirals with no evidence of a low luminosity turnover 
prior to M$_{K}$$-$5log$_{10}{\it h}~{<}$ $-$19.75\,mag. An exponential turnover is expected at the
high luminosity end.

Our findings with regard to the $K$-band LF are therefore threefold.
First, we find a total LF that is consistent with previous work. Second, each of the three galaxy classes (ellipticals, bulge-dominated spirals, 
and disk-dominated late type spirals) have a LF with a distinct shape, and
none of them mimics the shape of the total LF.  Third, ellipticals dominate the space density at high luminosities, a result that is only accentuated if total magnitudes are considered (section 4.4), whereas late-type (Sc - Scd) spirals dominate the space
density at low luminosities. Lying between these two extremes are the lenticular galaxies and the bulge-dominated spirals (S0/a - Sbc). 

The only other study to have explored the morphological type dependence of  $K$-band LFs is that of \cite{Koc01}. Using visual morphological classifications, they segregated their sample of bright, nearby galaxies ($K_{20} \le 11.25$ mag, cz $>$ 2000 km/s) into just two broad categories; early and late.  Figure~8 illustrates the agreement  between our results and those of  \cite{Koc01} when our sample is divided the same
way as theirs, at T = $-$0.5, such
that the early type sub-sample includes elliptical and lenticular galaxies and the late-type sub-sample
includes all galaxies classified S0/a and later. The shape of the LFs
are similar though not identical when galaxies are divided into these two broad categories. It is not until galaxies are more finely segregated that the differences between the LFs for the morphological  types emerge.

\subsection{$K$-band Luminosity Density by Hubble type}

Calculating the luminosity density is important as it can provide a constraint on the mass density of stars
in galaxies, given a mass to light ratio. 
Parameterizing the LFs allows the luminosity density {\it j} 
to be calculated by integration,

\begin{equation}
j =  \int \phi (M) 10^{0.4(M_\odot - M)} dM
\end{equation}

where M${_\odot}$ is the absolute magnitude of the Sun, corresponding to 3.32\,mag at $K$ \citep{Bel03}.
The total $K$-band luminosity density was calculated by integrating equation 2 over the
interval $-$25 ${\le }$ M$_{K}$$-$5log$_{10}{\it h}$ ${\le}$$-$19\,mag using the Schechter 
function illustrated in Figure~7.  This yields (5.8 ${\pm}$ 1.2) ${\times}$ 10$^{8}$ ${\it h}$ L$_{\odot}$ Mpc $^{-3}$. 
A Monte Carlo method was used to calculate the uncertainty.  Our value for the 
$K$-band total luminosity density agrees well with previous determinations 
\citep{Jon06,Bel03,Koc01,Col01}, all of which are about a factor of two lower than 
the value reported by \cite{Hua03}.  See \cite{Bel03} for a discussion of the various 
luminosity density estimates in the literature.

Elliptical galaxies (Figure~7) contribute ${\sim}$16 ${\pm}$ 3${\%}$ of the total ${K}$-band luminosity
density of galaxies in the local universe.  This value is revised
upwards to ${\sim}$18\% if total rather than isophotal magnitudes are considered (section 4.4).
Lenticulars and bulge-dominated spirals (Figure~7) contribute ${\sim}68 {\pm} 14{\%}$ 
of the total, or ${\sim}$22 ${\pm}$ 4${\%}$ for each sub-group (S0, S0/a--Sab, Sb--Sbc). 
Finally, the late-type spirals (Figure~7) contribute ${\sim}$16 ${\pm}$ 3${\%}$ of the 
total.  The results are summarized in Table~5.  {\it To a good approximation, 
each Hubble type (E, S0, S0/a--Sab, Sb--Sbc, Sc--Scd) contributes equally to the 
overall $K$-band luminosity density in the local universe.}  

\subsection{$K$-band Bulge Luminosity Functions by Hubble type}

In addition to the LFs for entire galaxies, the LF of bulges alone can provide information on how these
structural components formed. For lenticular and spiral galaxies, it is possible, in principle, to 
decompose each galaxy into its bulge and disk components by modeling imaging data \citep[e.g.,][]{Pen02,Sim02}.  An alternative approach, employed by \cite{Gra08}, is to compute the bulge and disk luminosities from S\'ersic bulge and exponential disk parameters. \cite{Gra08} have done this for a selection of about 400 nearby galaxies with ${K}$-band imaging in the published literature, 
many of which are in the K10/3000 sample. Thus, the resulting bulge/total luminosity ratios can be used statistically to predict bulge LFs for different Hubble types in the K10/3000 sample.

The total magnitudes adopted for spiral and lenticular galaxies are those reported in the 2MASS XSC
under the parameter: {\tt k${_{-}}$m${_{-}}$ext}. The total magnitudes were then corrected to bulge magnitudes
using the bulge/total luminosity ratios as a function of morphological type reported in \cite{Gra08}. 
Galaxies exhibit a range of bulge/total luminosity ratios even within a particular morphological type.
Following \cite{Gra08}, the adopted 1${\sigma}$ ranges for m$_{bulge}$ $-$ m$_{total}$ are 1.12 to 2.18 mag for the S0--S0/a  galaxies, 0.58 to 2.49 mag for the Sa--Sab galaxies, and 1.23 to 3.25 mag for the Sb--Sbc galaxies. A plausible bulge magnitude was estimated for each
galaxy using a Monte Carlo method that applied a type specific m$_{bulge}$ $-$ m$_{total}$ correction
to the total magnitude of each galaxy, randomly selected from a uniform distribution\footnotemark \footnotetext{ It is not clear which distribution to adopt because m$_{bulge}$ $-$ m$_{total}$ has been measured for only a few galaxies in the $K$-band. However, similar results are
obtained if a normal distribution is used instead of a uniform one. In either case, the mean correction 
affects the lateral displacement of the LF along the abscissa and the range of the correction affects the 
uncertainty which is reflected in the width of the band (Figure~9).} of real numbers in the ranges quoted above. A bulge LF was then computed using the Choloniewski method. This procedure was repeated 55 times leading to an average LF and an associated standard deviation. The mean bulge LF and the 1${\sigma}$ standard deviation from the mean is reported in Table 6 for each morphological type.

Ellipticals do not have disks and so there is no m$_{bulge}$ $-$ m$_{total}$ correction. However, the 2MASS isophotal magnitudes for the ellipticals underestimate the total magnitudes due to the fact that a restriction was imposed on the value of the S\'ersic index when the  2MASS total magnitudes were computed \citep{Lau07}. We determined the value 
m$_{iso}$ $-$ m$_{total}$ = 0.44 ${\pm}$ 0.28 mag by comparing the 2MASS XSC ${K}$-band isophotal magnitudes with the total ${K}$-band magnitudes measured by \cite{Mar03} using GALFIT \citep{Pen02} for 14 ellipticals in the K10/3000
sample. We found no dependence of m$_{iso}$ $-$ m$_{total}$ with absolute $K$-band magnitude.
Thus, total magnitudes are estimated by applying
a correction, randomly selected from a uniform distribution
of numbers in the range $-$0.16  to $-$0.72 mag, to the isophotal magnitude 
of each elliptical in the K10/3000 sample. Then a LF was computed using the Choloniewski method. This procedure was repeated 55 separate times leading to the average LF and the associated 
1${\sigma}$ standard deviation reported for ellipticals in Table 6.

The derived bulge LFs, illustrated in Figure~9,  
predict that the $K$-band bulge LF is bimodal with the ellipticals clearly offset from the bulges
of lenticulars and S0/a - Sbc spiral galaxies. This is a consequence of the type dependent {\em magnitude} corrections employed to obtain the bulge LFs. The
corrections essentially shift the isophotal LFs (Figure~7) towards {\em higher} luminosities for the ellipticals and towards {\em lower} luminosites for the lenticulars and the S0/a - Sbc spiral galaxies, thereby accentuating the difference between the types already noted in section 4.2.
Our combined bulge LF is similar in shape and amplitude 
to the ${B}$-band one reported by \cite{Dri07}, although a detailed comparison is complicated by the different wavelengths. The dwarf elliptical sequence {\em appears} to connect onto
the LFs for spiral bulges. However,  our LF for dwarf ellipticals is incomplete, these objects
have a distinct LF that continues to rise towards low luminosities as shown previously by \cite{San85}.

Integrating the bulge LFs using equation~3 allows an estimate of the contribution of bulges to the $K$-band luminosity density in the local universe. The results are presented in Table 7. Interestingly, 
bulges in each of the Hubble types, S0--S0/a, Sa--Sab, and Sb--Sbc contribute approximately equally
to the luminosity density. Ellipticals, on the other hand,  contribute about 50\% more to 
luminosity density than the lenticular and spiral bulges combined. Collectively, all bulges contribute 30 ${\pm}$ 7\% of the total $K$-band luminosity density of galaxies in the local universe, thus the remaining 70 ${\pm}$ 7\% is attributed to disks.  A similar result was obtained previously in the $V$-band by \cite{Sch87}.

\section{Discussion}

\subsection{Morphological Type Dependence of the $K$-band Luminosity Function}



Our principal new result is that the $K$-band isophotal LFs depend significantly on galaxy morphology. 
Broadly speaking, the $K$-band LFs for galaxies manifest in two varieties, both of which can be described 
mathematically by a Schechter function.  The LFs for ellipticals, lenticulars 
and bulge-dominated spirals (S0/a - Sbc) are peaked with a fall-off at high and low luminosities.
Although the functional forms for various types of {\it bulge} dominated galaxies are similar, the
ellipticals are displaced about one magnitude brighter in luminosity compared to the lenticulars
and bulge-dominated spirals, a difference that increases to ${\sim}$1.4 mag if total magnitudes are considered (section 4.4).
In contrast, the disk-dominated
late-type spirals (Sc--Scd) are distributed according to a power law with lower luminosity systems more numerous than any other galaxy type. 

Overall, our results agree qualitatively with \cite{Bin88} and \cite{Lov92}, who showed that the $B$-band LFs are different when galaxies are segregated by visible morphology. However, our results differ in detail, particularly with regard to the Sc galaxies. 
Subsequently, little evidence has been found for 
{\it type dependent} differences in the $B$--, SDSS-- and 2MASS--band LFs for morphological types earlier than Sc--Scd \citep{Efs88, Mar94, Mar98, Koc01, Nak03}. However, those galaxy samples were divided into two or at most three broad morphological classes whereas our results are based on a finer morphological segregation. Another factor to consider is the reliability of morphological assignments for distant galaxies.  Morphological
classification becomes difficult for
distant and faint galaxies, even more so when the morphological assignment is
based on photographic Schmidt
plates. As  noted previously by \citet{Lap03}, variations among the LFs  in the
literature depend largely on the criteria employed for typing the galaxies and the subsequent grouping of types.  When the different morphologies are bundled together the distinctions between the LFs become less apparent as Figure 8
illustrates.
Our results are based on well resolved nearby galaxies, so the classifications are more dependable, and 
the distinctions revealed by our analysis resulted from dividing the galaxy sample into several morphological bins.

Given the difficulties with morphological classifications for distant galaxies one is compelled to  critically evaluate the efficacy of surrogate but {\it quantitative} measures of galaxy morphology, such as the SDSS color and light concentration indices \citep{Str01}, both of which yield significant differences in the resulting LFs 
\cite[e.g.,][] {Bla01,Bel03,Con03, Bel04, Wil06, Bla06, Fab07}.
Finding the ideal combination of parameters
that can also be applied to galaxy simulations is one of the current observational challenges.
However subjective the visual classifications may be, their principal merit is that they do lead to distinct functional forms for the $K$-band LFs. 
Perhaps a useful compromise between the two approaches
would be to classify galaxies using artificial neural networks (ANN), although this method too seems to have problems with ANN classifications leading to similar Schechter functions for all galaxy types  \citep[e.g.,][]{Bal04,Bal06}. The Zurich Estimator of Structural Types (ZEST) is another approach 
introduced by \cite{Sca07} that appears to do better by returning different forms for the LFs of {\it bulge} dominated and {\it disk} dominated galaxies that more closely resemble the distinctions revealed by our analysis. 

The differing LFs for  {\it bulge} dominated and  {\it disk} 
dominated systems suggests at least two quite distinct galaxy formation mechanisms are at work to produce the diversity of morphological types seen in the local universe. The next step is to establish what the formation mechanisms are, which will require modeling the LFs in the context of hierarchical clustering scenarios \citep[e.g.,][]{Col00,Ben03}. Semi-analytic models have revealed that a combination
of cold gas accretion \citep{Wei04} and feedback \citep{Opp06} can yield a mass function for galaxy disks that is similar to, but slightly flatter than, the slope of the halo mass function. 
Such models are also able to reproduce the peaked LFs observed for ellipticals and bulge-dominated spirals by incorporating major mergers \citep{Bar92,Hop08b}. Work already underway shows that the morphological dichotomy revealed by the $K$-band LFs can be understood within the context of 
galaxy evolution in ${\Lambda}$CDM cosmology (A. Benson \& N. Devereux 2009, in preparation).
 
\subsection{The Hubble Type Dependence of Black Hole Mass Functions}

Most astronomers now believe that supermassive black holes (BHs) reside in the bulges of, possibly, all galaxies. The principal evidence is a correlation between the bulge luminosity, measured in the near-infrared, and black hole mass  \citep{Mar03, Fer00}. The importance of this correlation is that the near-infrared luminosity of stellar bulges, a component on which the morphological classification of galaxies is based, may be used as a surrogate tracer of the extragalactic black hole mass function (BHMF) which in turn constrains physical models of black hole growth in the context of  ${\Lambda}$CDM cosmology \citep{Mar04, Hop08a, Hop08b, Sha09}.

Using the group 1 calibration of \citet{Mar03}, the bulge LFs shown in Figure~9 may be translated into BHMFs.  
When defining BHMFs this way, it is important to appreciate the following caveats. Firstly, the group 1 calibration of \citet{Mar03} is based primarily on elliptical galaxies, thus one has to assume that the same relation applies to lenticulars and the bulges of spirals\footnotemark. \footnotetext{Sc and later types
have not been included in Figure 9 as central supermassive BHs have yet to be found in later type spirals \citep{Geb01}.}Secondly, it has not yet been proven that BHs 
exist in {\it all} ellipticals, lenticulars and the bulges of {\it all} spirals, thus the derived space densities will be upper limits. 
Nevertheless, if BHs do exist in all bulges, and the same bulge luminosity - BH mass correlation holds for all, then our results predict
that the BHMF is bimodal, and depends on Hubble type, with the ellipticals having a range of BH masses distinctly higher than the range of BH masses inhabiting lenticular and spiral galaxies.
Our total BHMF is similar to other total
BHMFs in the published literature which have been derived using a variety of relationships between BH mass
and bulge properties. As summarized by \citet{Sha09}, many of the existing BHMF determinations are based on total galaxy LFs, which include all morphological types, with subsequent corrections for the contribution of each morphological type to the total LF, and a further correction to estimate the contribution of the spheroid to the
luminosity of each type. Ours is the first to
be based on LFs explicitly segregated by morphological type. 
With the aforementioned caveats in mind, our results further predict that the majority of BHs reside
in lenticulars and the bulges of spirals. In contrast, the BHMF of \cite{Gra07}, which is based on an empirical non-linear relationship between BH mass and bulge S\'ersic parameter, predicts a sharp decrease in the BHMF over that same interval (Figure 9). Which of these two vastly different predictions is correct hinges on whether the bulge luminosity \citep[e.g.,][]{Mar03} or the bulge S\'ersic parameter \citep[e.g.,][and references therein]{Gra07} is the better predictor of the existence and mass of central BHs in lenticulars and the bulges of spiral galaxies. 

Given the aforementioned caveats, an upper limit to the local mass density of BHs, ${\rho_{\bullet}}$, may be calculated by integrating the BHMFs using

\begin{equation}
\rho_{\bullet} =  \int^{M_{upper}}_{M_{lower}} \phi (M_{\bullet}) M_{\bullet}  d(log M_{\bullet})
\end{equation}

where ${\phi (M_{\bullet})}$ is the differential BH mass function, mapped from the bulge LF
(Figure~9), and M${_{\bullet}}$ is the BH mass.

The result of the integral is very  sensitive to the value of the upper mass limit which undoubtedly
contributes to the
factor of ${\sim}$2 range in estimates for ${\rho_{\bullet}}$ in the literature. Integrating between 6.5 ${\le}$ log$_{10}$(M$_{BH}$) ${\le}$ 9.4, where the upper mass limit corresponds to the highest {\em measured} BH mass \citep{Kas07} yields a total BH mass density of (4.6 ${\pm}$ 0.5) x 10$^{5}$ M${_\odot}$ Mpc$^{-3}$ ($h$ = 0.7) in good agreement with previous determinations \citep[][and references therein]{Sha09}. However, our prediction that the BHMF is bimodal
associates over half, (${\sim}$ 60\%), of the total black hole mass density with elliptical galaxies, the bulges of spirals making up the remainder, apportioned approximately equally between the lenticulars, the S0/a-Sab's and the Sb-Sbc's. 

At present, stellar bulges provide an efficient but indirect means of constraining
supermassive black hole demographics without recourse to the difficult and time-consuming 
measurement of black hole masses {\it directly} using gas and star kinematics. In particular, 
because stellar bulge luminosities can be measured consistently 
for very large numbers of galaxies, the method has the potential to 
reduce biases that may affect other {\it indirect} methods, such as the use of broad emission lines \citep[e.g.][]{Gre07}. We are currently undertaking the next logical step to refine this analysis which
is to isolate the bulges explicitly using the 2MASS images.





 \section{Conclusions}

$K$-band LFs have been presented for a complete volume limited sample of 1613 galaxies with V$_{\mathrm gsr}$ ${\le}$ 3000\,km\,s$^{-1}$, K  ${\le}$ 10 mag and ${|b|}$  ${>}$ 10 degrees. Our principal 
conclusions are \\
\\
1.  The $K$-band LFs depend on morphological type when galaxies are segregated according to the visual classification scheme of \cite{deV91}. Ellipticals dominate the space density at high luminosities whereas late-type spirals dominate 
the space density at low luminosities. Lying in between the two extremes are the lenticulars and bulge-dominated spirals.  
\\
\\
2. The $K$-band LF for late-type spirals follows a power law that rises towards low luminosities whereas the $K$-band
LFs for ellipticals, lenticulars and bulge-dominated spirals are peaked with a falloff at both high and low luminosities.
\\
\\
3. Each morphological type (E, S0, S0/a--Sab, Sb--Sbc, Sc--Scd) contributes approximately equally to the overall $K$-band luminosity density in the local universe.
\\
\\
4. The $K$-band LF of bulges is bimodal and depends on morphological type. 
Ellipticals comprise 60\% of the bulge luminosity density in the local universe,
the remaining 40\%  is associated with lenticulars and the bulges of spirals.
Overall, bulges contribute ${\sim}$ 30\% of the total galaxy luminosity density at $K$, with disks making up the difference.
\\
\\
5. If the bulge luminosity traces the black hole mass, our results predict
that the black hole mass function is bimodal and depends on morphological type
with the most massive black holes occurring in ellipticals and the space density of black holes
reaching a maximum in the bulges of spirals.




\acknowledgments

N.D. gratefully acknowledges the support of the Fulbright Commission and the hospitality of the National University of Ireland, Galway, and the assistance of 
undergraduate students Meghan Burleigh, Sarah McNamara and Lisa Fallon for their help in the early stages of this project. Extensive use was made of the HyperLeda database (http://leda.univ-lyon1.fr) and data products from the Two Micron All Sky Survey, which is a joint project of the University of Massachusetts and the Infrared Processing and Analysis Center/California Institute of Technology, funded by the National Aeronautics and Space Administration and the National Science Foundation.
The Millennium Simulation databases used in this paper and the web
application providing online access to them were constructed as part
of the activities of the German Astrophysical Virtual Observatory.
We thank Darren Croton and Gerard Lemson for help with the
database tools. We thank Daniel Eisenstein for
supplying the galaxy correlation function used in the cosmic variance
calculation. We thank John Huchra for providing galaxy redshifts in advance of publication
and Alessandro Marconi and Francesco Shankar for digitized BHMFs.



{\it Facilities:}  \facility{2MASS}



\clearpage

\appendix

\section{Appendix}

This Appendix describes in detail how to calculate the size of
the error bars on the differential LF $ {\phi}$(M), the number density of galaxies 
${\rho}$(${\mu}$), and n, the average number density, where M is absolute magnitude and  ${\mu}$ is the distance modulus. The reader is encouraged to consult
\cite{Cho86} prior to implementing the following strategy which requires a familiarity with the maximum-likelihood method of statistical analysis.

In order to characterize the statistical uncertainties on $ {\phi}$(M),  ${\rho}$(${\mu}$), and n, the 
observables, M and ${\mu}$, must be represented by a particular statistical model. Generally, this model is ${\it assumed}$ to be a Poisson distribution, which means that the galaxies are distributed independently of position, that the average number of galaxies in a region of space is proportional only to the volume of that region, and that their luminosities are uncorrelated with their locations.  This model yields the familiar
and widely adopted result 1/(2.3${\sqrt N}$) for the uncertainty in log$_{10}$${\phi}$(M), where N
is the number of galaxies in the absolute magnitude interval, ${\Delta}$M. However,
it is rarely, if ever, proven that the observables, M and ${\mu}$, {\it are} distributed in a Poisson
fashion even though there are reasons to believe that this may not be  a good assumption given the clustering
tendencies for elliptical galaxies in particular. We have discovered, in the course of writing this paper,
that some uncertainties calculated using the procedure described below are less than 1/2.3${\sqrt N}$. These cases are indicated by asterisks in Table 3. This outcome 
suggests that the galaxy distribution is not Poisson, but is instead slightly clumped in space and luminosity and better represented by a \textit{generalized Poisson distribution} \citep{Con89}.  
The  \textit{generalized Poisson} distribution is described by 
\begin{equation}
p(N|V,b)=\frac{\bar{N}(1-b)}{N!}[\bar{N}(1-b)+N b]^{N-1}e^{-\bar{N}(1-b)-Nb}.
\end{equation}
Here $p(N|V,b)$ is the probability that a cell of volume $V$ placed randomly in space contains exactly $N$ galaxies. If n is the average density of galaxies then ${\bar{N}}$ = n${V.}$ An application of this distribution to hierarchical clustering is described by \cite{She98} and \cite{Sas84}. A formal analysis, similar to that outlined below, but beyond the scope of the present paper, would constrain the parameter $b$ to the range $0 \le b \le$ 1 with the standard Poisson distribution corresponding to the case $b=0$. Such a distribution would yield uncertainties that
are larger than 1/2.3${\sqrt N}$ by a factor $1/(1- b)$.

Following \cite{Cho86}, we proceed with the Poisson distribution function, and adopt the
notation of  \cite{Cho86}, which is most easily understood with
reference to Figure 2 of that paper. ${A}$ is the total number of absolute magnitude columns, indexed by the
counter ${ i}$, and ${B}$ is the total number of distance modulus rows, indexed by the counter ${j}$.  A magnitude limited sample fills the M, ${\mu}$ plane only partially, in the region below the selection line M + ${\mu}$ = m$_{lim}$, where m$_{lim}$ is the magnitude limit of the survey. The parameter S, is the 
magnitude limit of the survey, m$_{lim}$, in units of  ${ i}$ and ${j}$, such that ${ i}$ + ${j}$ ${\le}$ S, 
an inequality because the galaxies are also contained within a finite volume and within a finite range of
absolute magnitudes.

The covariance, a matrix whose leading diagonal contains the uncertainties for the  values, $\phi_i,\rho_j, n$,
can be approximated by use of the information matrix, specifically, 
\begin{equation}
Cov(\hat{E}) \approx \left[- \frac{\partial^2 ln(l)}{\partial E^2}\right] ^{-1}_{E=\hat{E}}
\end{equation}
where ${l}$ is the likelihood function, which is chosen to be Poisson. The minus sign corrects a typo in \cite{Cho86}, \citep[see][]{Ead82}. Following \cite{Cho86},  $E = (\phi_i,\rho_j,n)$, are called the ${estimators}$
and are the computed values defining $ {\phi}$(M),  the differential LF, where M is absolute magnitude, 
${\rho}$(${\mu}$), the number density of galaxies as a function of distance modulus, ${\mu}$, and n, the average number density of galaxies in the sample.

The Hessian $\frac{\partial^2 ln(l)}{\partial E^2}$ of $l$ can be written as a 3 by 3 symmetric matrix with each entry itself a matrix:

\begin{equation} \frac{\partial^2 ln(l)}{\partial E^2} =
\left[ \begin{array}{ccc}
\frac{\partial^2 ln(l)}{\partial \phi^2} & \frac{\partial^2 ln(l)}{\partial \phi \partial \rho} & \frac{\partial^2 ln(l)}{\partial \phi \partial n} \\
\frac{\partial^2 ln(l)}{\partial \rho \partial \phi} & \frac{\partial^2 ln(l)}{\partial \rho ^2} & \frac{\partial^2 ln(l)}{\partial \rho \partial n} \\

\frac{\partial^2 ln(l)}{\partial n \partial \phi} & \frac{\partial^2 ln(l)}{\partial n \partial \rho} & \frac{\partial^2 ln(l)}{ \partial n^2}  \\
\end{array} \right] .
\end{equation} 
So, the next step in the procedure is clear, but complicated, because it entails taking the second derivative of the likelihood function with respect to each of the variables, ${\phi_i, \rho_j}$ and ${n}$.

The Poisson likelihood function is given by
\begin{equation}
	l=\begin{array}{c}
	{\prod_{i=1}^A \prod_{j=1}^B} \\
	\null_{ i+j \leq S}  \end{array}
	  e^{-\lambda_{i,j}}\frac{\lambda_{i,j}^{N_{i,j}}}{N_{i,j}!} 
\end{equation}
where $\lambda_{i, j} = \frac{1}{n} \phi_{i} \Delta M \rho_{j } ( V_{j} -V_{j -1})$,
and $V$ is the sample volume and $\Delta M$ is the magnitude interval. 

Taking the natural log we obtain,
\begin{equation} 
ln(l)=\begin{array}{c}
	{\sum_{i=1}^A \sum_{j=1}^B} \\
	\null_{ i+j \leq S}  \end{array}
	  e^{-\lambda_{i,j}}\frac{\lambda_{i,j}^{N_{i,j}}}{N_{i,j}!}  -\lambda_{i,j} +N_{i,j} ln(\lambda_{i,j})-ln(N_{i,j}!) 
\end{equation}
Unless otherwise stated double products such as $ 
	\begin{array}{c}
	{\prod_{i=1}^A \prod_{j=1}^B} \\
	\null_{ i+j \leq S}  \end{array} $ 
 will be denoted by $\prod_{i+j \leq S}$
 and double sums such as $ 
	\begin{array}{c}
	{\sum_{i=1}^A \sum_{j=1}^B} \\
	\null_{ i+j \leq S}  \end{array}$
will be denoted by $\sum_{i+j \leq S}$.

Here $N_{i,j} =$ the number of galaxies in bin i,j. Consequently 
\begin{equation} \frac{\partial N_{i,j}}{\partial \phi_k}=\frac{\partial N_{i,j}}{\partial \rho_k}= \frac{\partial N_{i,j}}{\partial n}=0. \end{equation}

For the first sub-matrix,
\begin{equation} \frac{\partial^2 ln(l)}{\partial \phi ^2} =
\left[ \begin{array}{ccc}
\frac{\partial^2 ln(l)}{\partial \phi_1 \partial \phi_1} & \cdots & \cdots \\
\vdots & \frac{\partial^2 ln(l)}{\partial \phi_i \partial \phi_j} & \vdots \\
\vdots & \cdots & \ddots  \\
\end{array} \right] .
\end{equation}

We have
\begin{equation}
 \frac{\partial^2 ln(l)}{\partial \phi_i \partial \phi_j}=\frac{\partial^2}{\partial \phi_i \partial \phi_j}\left( \sum_{\alpha + \beta \leq S}-\lambda_{\alpha \beta}+N_{\alpha \beta}ln(\lambda_{\alpha \beta})-ln(N_{\alpha \beta}!)\right).
\end{equation}
By definition $\lambda_{\alpha \beta} = \frac{1}{n} \phi_{\alpha} \Delta M \rho_{\beta } ( V_{\beta} -V_{\beta -1})$,
where $V$ is the sample volume and $\Delta M$ is the magnitude interval. We introduce new dummy variables $\alpha,\beta$ which are independent of the variables $i,j$. This is necessitated by the need to distinguish between the variables which the likelihood is a function of and the factors which the likelihood is composed of.

Letting $C_{\beta} =\Delta M(V_{\beta}-V_{\beta -1})$, this becomes 
$\lambda_{\alpha \beta}=C_{\beta}\frac{1}{n}\phi_{\alpha}\rho_{\beta},$ and so
 
\begin{equation}
\frac{\partial \lambda_{\alpha \beta}}{\partial \phi_i}= 
\begin{array}{ll}
 0 & \mbox{if $\quad \alpha \ne i $,}\\
C_{\beta} \frac{1}{n} \rho_{\beta} &\mbox{if $\quad \alpha = i $.}
\end{array}
\end{equation}
So 
$\frac{\partial^2 \lambda_{\alpha \beta}}{\partial \phi_i \partial \phi_j}=0$  for all $\alpha, \beta, i,j$.
Also 
$\frac{\partial^2 N_{\alpha \beta}!}{\partial \phi_i \partial \phi_j}=0$  for all $\alpha, \beta, i,j$. Consequently 
\begin{equation}
 \frac{\partial^2 ln(l)}{\partial \phi_i \partial \phi_j}= \frac{\partial^2 }{\partial \phi_i \partial \phi_j}\left( \sum_{\alpha + \beta \leq S}N_{\alpha \beta} ln(\lambda_{\alpha \beta} )\right).
\end{equation}
Now
\begin{equation}
\frac{\partial }{ \partial \phi_j}\left( \sum_{\alpha + \beta \leq S}N_{\alpha \beta} ln(\lambda_{\alpha \beta} )\right)=
 \sum_{\alpha + \beta \leq S}N_{\alpha \beta} \frac{\partial ln(\lambda_{\alpha \beta})}{\partial  \phi_j}.
\end{equation}
\begin{equation}
=
 \sum_{\alpha + \beta \leq S}N_{\alpha \beta} \frac{\partial ln(C_{\beta}\frac{1}{n} \phi_{\alpha} \rho_{\beta})}{\partial  \phi_j}
\end{equation}
\begin{equation}
=
 \sum_{ \beta =1}^{ S -j}N_{j \beta} \frac{\partial ln(C_{\beta}\frac{1}{n} \phi_j \rho_{\beta})}{\partial  \phi_j}
=
 \sum_{ \beta =1}^{ S -j}N_{j \beta} \frac{1}{\phi_j}.
\end{equation}
So 
\begin{equation}
 \frac{\partial^2 ln(l)}{\partial \phi_i \partial \phi_j}= \frac{\partial }{\partial \phi_i}\sum_{ \beta =1}^{ S -j}N_{j \beta} \frac{1}{\phi_j}
\end{equation}
\begin{equation}
=\sum_{ \beta =1}^{ S -j}N_{j \beta} \frac{\partial (\phi_j^{-1})}{\partial \phi_i}
\end{equation}

\begin{equation} 
=
\begin{array}{ll}
 \frac{-1}{\phi_i^2}\sum_{ \beta =1}^{ S -i}N_{i \beta} & \mbox{if $i=j $,}
\\
0 &\mbox{if $i \ne j $}
\end{array}
\end{equation}
Replacing the dummy variable $\beta$ by the variable $j$ we obtain
\begin{equation}
\frac{\partial^2 ln(l)}{\partial \phi_i \partial \phi_j}=
\begin{array}{ll}  \frac{-1}{\phi_i^2}\sum_{ j =1}^{ S -i}N_{i j} & \mbox{if $i=j $,}
\\
0 &\mbox{if $i \ne j $}
\end{array}
\end{equation}
Therefore the first block of the Hessian matrix is an ${A}$ x ${A}$ matrix
\begin{equation}
\left[ \begin{array}{c}
\frac{\partial^2 ln(l)}{\partial \phi^2 } \\
\end{array} \right] =
\left[ \begin{array}{ccccc}
\frac{-\sum_{ j =1}^{ S -1}N_{1 j}}{\phi_i^2}&0&0\\
0&\frac{-\sum_{ j =1}^{ S -2}N_{2 j}}{\phi_i^2}&0&\\
0&0&\ddots&\\
\end{array} \right].
\end{equation}
in which all terms are zero, except the leading diagonal, which is just the sum of the galaxies in each column, ${i}$, divided by -${\phi_i^2}$.

We next calculate $\frac{\partial^2 ln(l)}{\partial \rho_i \partial \rho_j}.$  
Since $\lambda_{\alpha \beta} =C_{\beta}\frac{1}{n}\phi_{\alpha}\rho_{\beta}$,
\begin{equation}
\frac{\partial \lambda_{\alpha \beta}}{\partial \rho_j}= 
\begin{array}{ll} 0 & \mbox{if $\quad \beta \ne j $,}
\\
C_j \frac{1}{n} \phi_{\alpha} &\mbox{if $\quad \beta = j $.}
\end{array}
\end{equation}
Therefore $\frac{\partial^2 \lambda_{\alpha \beta}}{\partial \rho_i \partial \rho_j}=0 $ for all $i,j,\alpha,\beta$ and 
\begin{equation}
\frac{\partial^2 ln(l)}{\partial \rho_i \partial \rho_j}=\frac{\partial^2 }{\partial \rho_i \partial \rho_j}\sum_{\alpha+\beta \le S} N_{\alpha \beta} ln(\lambda_{\alpha \beta}).
\end{equation}
One has \begin{equation}\frac{\partial }{\partial \rho_j}\sum_{\alpha+\beta \le S} N_{\alpha \beta} ln(\lambda_{\alpha \beta}) = 
\sum_{\alpha+\beta \le S} N_{\alpha \beta} \frac{\partial}{\partial \rho_j}ln(\lambda_{\alpha \beta})=
\sum_{\alpha=1}^ { S-j} N_{\alpha j} \frac{\partial}{\partial \rho_j}ln(C_\beta \frac{1}{n}\phi_{\alpha}\rho_j)
\end{equation}
\begin{equation}=
\sum_{\alpha=1}^ { S-j} N_{\alpha j} \frac{\partial}{\partial \rho_j}(ln(C_\beta \frac{1}{n}\phi_{\alpha})+ln(\rho_j))=\sum_{\alpha=1}^ { S-j}  \frac{N_{\alpha j}}{\rho_j}.
\end{equation}
Therefore
\begin{equation}
\frac{\partial^2 ln(l)}{\partial \rho_i \partial \rho_j}=\frac{\partial}{\partial \rho_i}\sum_{\alpha=1}^ { S-j}  \frac{N_{\alpha j}}{\rho_j}
\end{equation}
\begin{equation} 
=\begin{array}{ll} 0 & \mbox{if $i \ne j $,}
\\
\sum_{\alpha=1}^ { S-j}(\frac{-N_{\alpha j}}{\rho_j^2}) &\mbox{if $ i = j $.}
\end{array}
\end{equation}
\begin{equation}
= 
\begin{array}{ll} 0 & \mbox{if $i \ne j $,}
\\
\frac{-\sum_{i=1}^ { S-j}N_{ij}}{\rho_j^2} &\mbox{if $ i = j $.}
\end{array}
\end{equation}
To summarize, the next block of the Hessian matrix is a ${B}$ x ${B}$ matrix
\begin{equation}
\left[ \begin{array}{c}
\frac{\partial^2 ln(l)}{\partial \rho \partial \rho} \\
\end{array} \right] =
\left[ \begin{array}{ccccc}
-\frac{\sum_{ i=1}^{ S -1 }N_{i 1}}{\rho_1^2}&0&0&\\
0&\frac{-\sum_{ i=1}^{  S -2 }N_{i 2}}{\rho_2^2}&0&\\
0&0&\ddots&\\
\end{array} \right].
\end{equation}
in which all terms are zero, except the leading diagonal, which is just the sum of the galaxies in each row, ${j}$, divided by -${\rho_j^2}$.

To calculate $\frac{\partial^2 ln(l)}{\partial n^2}$ we proceed as follows:
Since $\lambda_{\alpha \beta} =C_{\beta} \frac{1}{n} \phi_{\alpha} \rho_{\beta}$, 
\begin{equation}
\frac{\partial \lambda_{\alpha \beta}}{\partial n} =-C_{\beta} n^{-2} \phi_{\alpha} \rho_{\beta},
\end{equation}
hence
\begin{equation}
\frac{\partial ^2 \lambda_{\alpha \beta}}{\partial n^2} =2C_{\beta} n^{-3} \phi_{\alpha} \rho_{\beta},
\end{equation}
and
\begin{equation}
\frac{\partial ^2 }{\partial n^2}\sum_{\alpha+\beta \leq S}-\lambda_{\alpha \beta} =-2\left( \sum_{\alpha+\beta \leq S} C_{\beta}\phi_{\alpha} \rho_{\beta}\right)  n^{-3} .
\end{equation}
Also
\begin{equation}
\frac{\partial ^2 ln(\lambda_{\alpha \beta})}{\partial n^2} =\frac{\partial ^2}{\partial n^2}\left( ln (C_{\beta} \phi_{\alpha} \rho_{\beta})-ln( n)\right) =-\frac{\partial ^2}{\partial n^2}\left( ln( n)\right) =n^{-2}.
\end{equation}
Consequently
\begin{equation}
\frac{\partial^2 ln(l)}{\partial n^2}=-2\left( \sum_{\alpha + \beta \leq S} C_{\beta} \phi_{\alpha}\rho_{\beta}\right) n^{-3}
+\left( \sum_{\alpha + \beta \leq S} N_{\alpha \beta}\right) n^{-2}
\end{equation}
\begin{equation}
=-2\frac{ \sum_{i+j \leq S} C_j \phi_i \rho_j}{ n^3}+\frac{\sum_{i+j \leq S} N_{ij}} {n^{2}}.
\end{equation}
a single scalar term.

We next calculate the first of the cross terms, $\frac{\partial^2 ln(l)}{\partial \phi \partial \rho}$. 

Since $\lambda_{\alpha \beta} =C_{\beta} \frac{1}{n} \phi_{\alpha} \rho_{\beta}$, 
\begin{equation}
\frac{\partial \lambda_{\alpha \beta}}{\partial \rho_j} =
\begin{array}{ll} 0 & \mbox{if $ \beta \ne j $,}
\\
C_j \frac{1}{n}\phi_{\alpha} &\mbox{if $ \beta = j $.}
\end{array}
\end{equation}
then 
\begin{equation}
\frac{\partial^2 \lambda_{\alpha \beta}}{\partial \phi_i \partial \rho_j} =
\begin{array}{ll} 0 & \mbox{if $ \beta \ne j $,}
\\
\frac{\partial} {\partial \phi_i} C_j \frac{1}{n}\phi_{\alpha} &\mbox{if $ \beta = j $.}
\end{array}
\end{equation}
\begin{equation}
 =
\begin{array}{ll} 0 & \mbox{if $ \beta \ne j$ or $\alpha \ne i $,}
\\
C_j \frac{1}{n} &\mbox{if $ \beta = j $ and $\alpha = i$.}
\end{array}
\end{equation}
Since 
\begin{equation}
\frac{\partial^2 ln(\lambda_{\alpha \beta})}{\partial \phi_i \partial \rho_j}=
\frac{\partial^2 }{\partial \phi_i \partial \rho_j}\left( ln C_{\beta}-ln (n) + ln \phi_{\alpha}+ln \rho_{\beta}\right) 
=
\frac{\partial }{\partial \phi_i }\left( 0-0 + \frac{\partial ln \phi_{\alpha}}{\partial \rho_j}+\frac{ln \rho_b}{\partial \rho_j}\right)=0.
\end{equation}
Consequently $\frac{\partial^2 ln(l)}{\partial \phi_i \partial \rho_j}= -\sum_{\alpha + \beta \leq S}
\frac{\partial^2 \lambda_{\alpha \beta}}{\partial \phi_i \partial \rho_j} $
\begin{equation}
 =
\begin{array}{ll}  \frac{C_j}{n} & \mbox{if $ i+j \leq S $,}
\\
0 &\mbox{otherwise.}
\end{array}
\end{equation}
Therefore:
\begin{equation} \frac{\partial^2 ln(l)}{\partial \phi \partial \rho} =
\left[ \begin{array}{cccccc}
-C_1/n & -C_1/n & -C_1/n& \cdots &C_1/n&C_1/n\\
-C_2/n & -C_2/n & -C_2/n & \cdots &C_2/n&C_2/n\\
-C_3/n & -C_3/n & -C_3/n & \cdots &C_3/n&C_3/n\\
-C_4/n & -C_4/n & -C_4/n& \cdots &C_4/n&0\\
-C_5/n & -C_5/n &-C_5/n& \cdots&0&0\\
\cdots & \cdots & \cdots&0&0&0\\
\cdots & \cdots&0 &0&0&0\\
-C_{B}/n & \cdots &0&0&0&0\\
\end{array} \right] .
\end{equation}
a {B} x {A} array, and its reflection, $\frac{\partial^2 ln(l)}{\partial \rho \partial \phi}$,  an {A} x {B} array.

To calculate the next cross term $ \frac{\partial^2 ln(l)}{\partial \phi \partial n}$:
\begin{equation}
\frac{\partial \lambda_{\alpha \beta}}{\partial n} =-C_{\beta} n^{-2} \phi_{\alpha} \rho_{\beta},
\end{equation}
so
\begin{equation}
\frac{\partial^2 \lambda_{\alpha \beta}}{\partial \phi_i \partial n}= 
\begin{array}{ll} 0 & \mbox{if $\alpha \ne i $,}
\\
\frac{-C_{\beta} \rho_{\beta}}{n^2} &\mbox{if $ \alpha = i $.}
\end{array}
\end{equation}
Therefore
\begin{equation}
\frac{\partial ^2 }{\partial \phi_i \partial n}\begin{array}{c}
	{\sum_{\alpha =1}^A \sum_{\beta =1}^B} \\
	\null_{ \alpha + \beta \leq S}  \end{array}\lambda_{\alpha \beta}=
\sum_{\beta=1}^ {B\wedge (S-i)} \frac{\partial^2 \lambda_{\alpha \beta}}{\partial \phi_i \partial n}=
-\frac{\sum_{\beta=1}^{ B\wedge (S-i)} C_{\beta}\rho_{\beta}}{n^2}.
\end{equation}

(where $x \wedge y $ denotes $min(x,y)$).

\begin{equation}
\frac{\partial^2 ln\lambda_{ij}}{\partial \phi_i \partial n} =
\frac{\partial ^2}{\partial \phi_i \partial n} \left( ln C_{\beta} -ln(n) +ln \phi _{\alpha} + ln \rho _{\beta}\right) 
=0-\frac{\partial}{\partial \phi_i}\frac{1}{n} +0+0=0.
\end{equation}
So 
\begin{equation}
\frac{\partial^2 ln(l)}{\partial \phi_i \partial n} =\frac{\sum_{j=1}^ { B\wedge (S-i)} C_{j}\rho_{j}}{n^2}
\end{equation}
and 
\begin{equation}\frac{\partial^2 ln(l)}{\partial \phi \partial n} =\left( \frac{\sum_{j=1}^{ B\wedge (S-1)} C_{j}\rho_{j}}{n^2},\frac{\sum_{j=1}^{B\wedge (S-2)} C_{j}\rho_{j}}{n^2}
\cdots,\frac{\sum_{j=1}^{ B\wedge (S-A)} C_{j}\rho_{j}}{n^2}\right)\end{equation}
a column vector, and its reflection, $ \frac{\partial^2 ln(l)}{\partial n \partial \phi}$, a row vector, each containing A terms.

Now, the final cross term,
\begin{equation}
\frac{\partial^2 ln(l)}{\partial \rho_j \partial n} 
\end{equation}
Since 
\begin{equation}
\frac{\partial \lambda_{\alpha \beta}}{\partial n} =-C_{\beta} n^{-2} \phi_{\alpha} \rho_{\beta},
\end{equation}
\begin{equation}
\frac{\partial^2 \lambda_{\alpha \beta}}{\partial \rho_j \partial n} =
\begin{array}{ll} 0 & \mbox{if $ \beta \ne j $,}
\\
\frac{-C_{\beta}\phi_{\alpha}}{n^2} &\mbox{if $ \beta = j $.}
\end{array}
\end{equation}

So
\begin{equation}
\frac{\partial^2 }{\partial \rho_j \partial n}\begin{array}{c}
	{\sum_{\alpha=1}^A \sum_{\beta=1}^B} \\
	\null_{ \alpha + \beta \leq S}  \end{array}\lambda_{\alpha \beta}=-\frac{\sum_{\alpha =1}^{ A \wedge (S-j)} C_j \phi_{\alpha}}{n^2}=-C_j\frac{\sum_{\alpha =1}^{ A \wedge (S-j)} \phi_{\alpha}}{n^2}=-C_j\frac{\sum_{i =1}^{ A \wedge (S-j)} \phi_i}{n^2}.
\end{equation}
To calculate $\frac{\partial^2 ln \lambda_{\alpha \beta}}{\partial \rho_j \partial n}$:
\begin{equation}
\frac{\partial ln \lambda_{\alpha \beta}}{\partial n} =\frac{\partial}{\partial n} \left( ln C_{\beta}- ln(n )+ ln \phi_{\alpha} + ln \rho_{\beta}\right) =-\frac{1}{n}.
\end{equation}
So
\begin{equation}
\frac{\partial^2 ln(\lambda_{\alpha \beta})}{\partial \rho_j \partial n} =-\frac{\partial 1/n}{\partial \rho_j}=0.
\end{equation}
Thus, 
\begin{equation} 
\frac{\partial^2 ln(l)}{\partial \rho_j \partial n} = C_j \frac{\sum_{i=1}^{ S-j}\phi_i}{n^2}.
\end{equation}
So, the last block of the Hessian matrix is,
\begin{equation}
\left( C_1 \frac{\sum_{i=1}^{ A \wedge (S-1)}\phi_i}{n^2},...,C_{B} \frac{\sum_{i=1}^{ A \wedge (S-B)}\phi_i}{n^2}\right) 
\end{equation}
another column vector, and its reflection,  $ \frac{\partial^2 ln(l)}{\partial n \partial \rho}$, a row vector, each containing B terms.

Finally, take the inverse of the negative of the (${A}$ + ${B}$ + 1) x (${A}$ + ${B}$ + 1) Hessian matrix, the leading diagonal of which contains the uncertainties on $ {\phi}$(M),  ${\rho}$(${\mu}$), and n.

\begin{figure}
\epsscale{1.0}
\begin{center}
\plotone{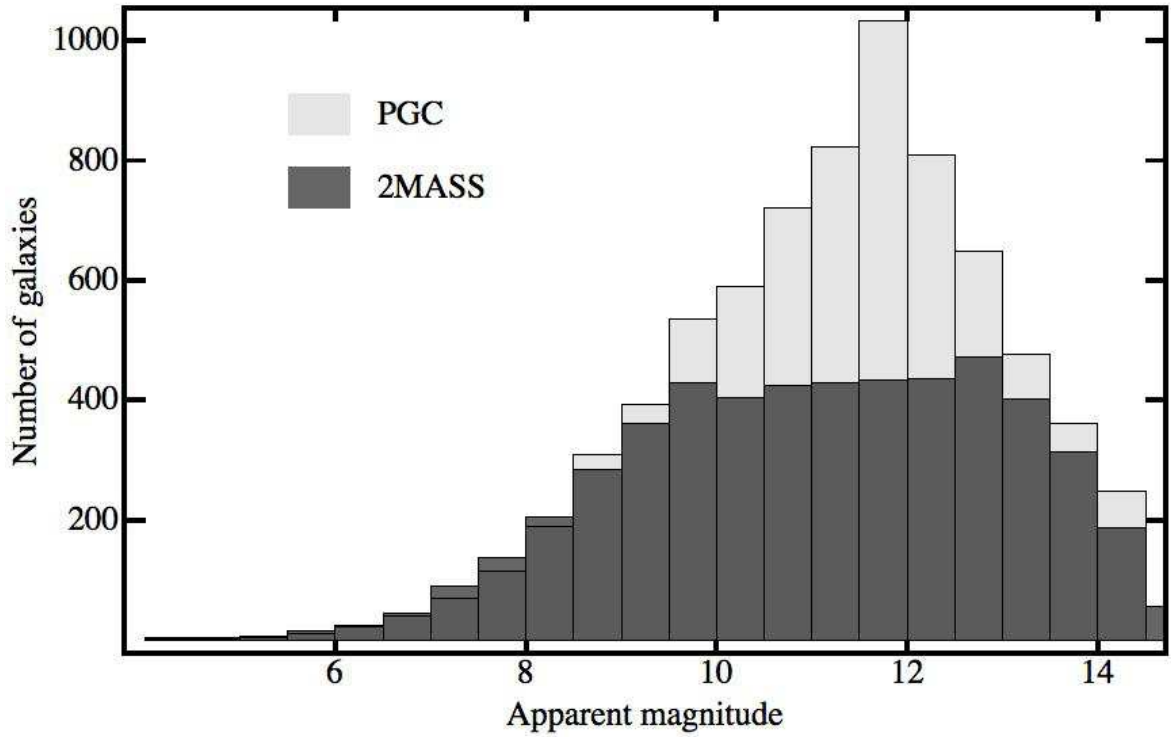}
\caption{ Histograms showing the distribution of B $-3.5$ magnitudes for a sample of 7406 nearby PGC galaxies (V$_{\mathrm gsr}$ ${\le}$ 3000\,km\,s$^{-1}$, ${|b|}$  ${>}$ 10$^{o}$, m$_{B}$ ${\le}$ 18 mag) and the distribution of $K$-band magnitudes for the subset of 5034 in the same volume that appear in the 2MASS XSC, and from which the K10/3000 galaxy sample is selected. }
\label{default}
\end{center}
\end{figure}

\begin{figure}
\epsscale{1.0}
\begin{center}
\plotone{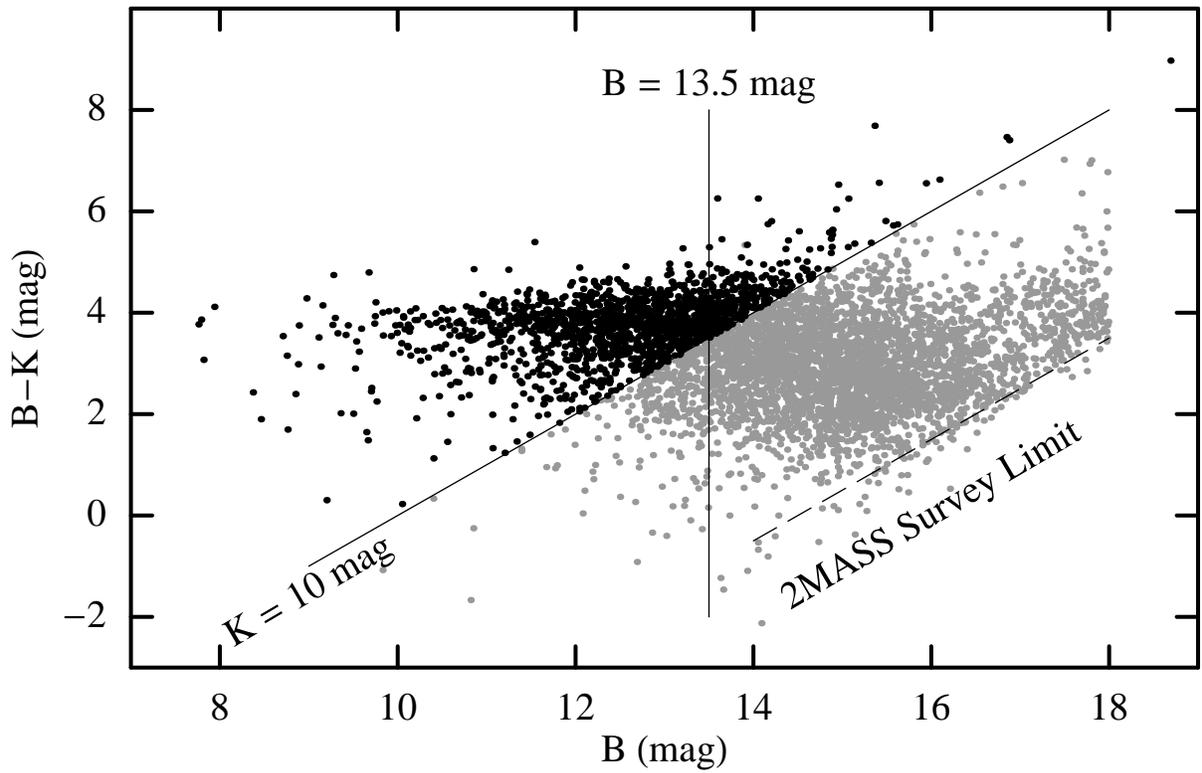}
\caption{B$-$K color versus apparent B magnitude for 5034 nearby PGC galaxies in the 2MASS XSC (V$_{\mathrm gsr}$ ${\le}$ 3000\,km\,s$^{-1}$, ${|b|}$  ${>}$ 10$^{o}$, m$_{B}$ ${\le}$ 18 mag). Black dots highlight 1596 galaxies in the 2MASS XSC with K ${\le}$ 10 mag that are contained within the same volume. A dashed line identifies the nominal 2MASS XSC limit at K = 14.5 mag.}
\label{default}
\end{center}
\end{figure}

\begin{figure}
\epsscale{1.0}
\begin{center}
\plotone{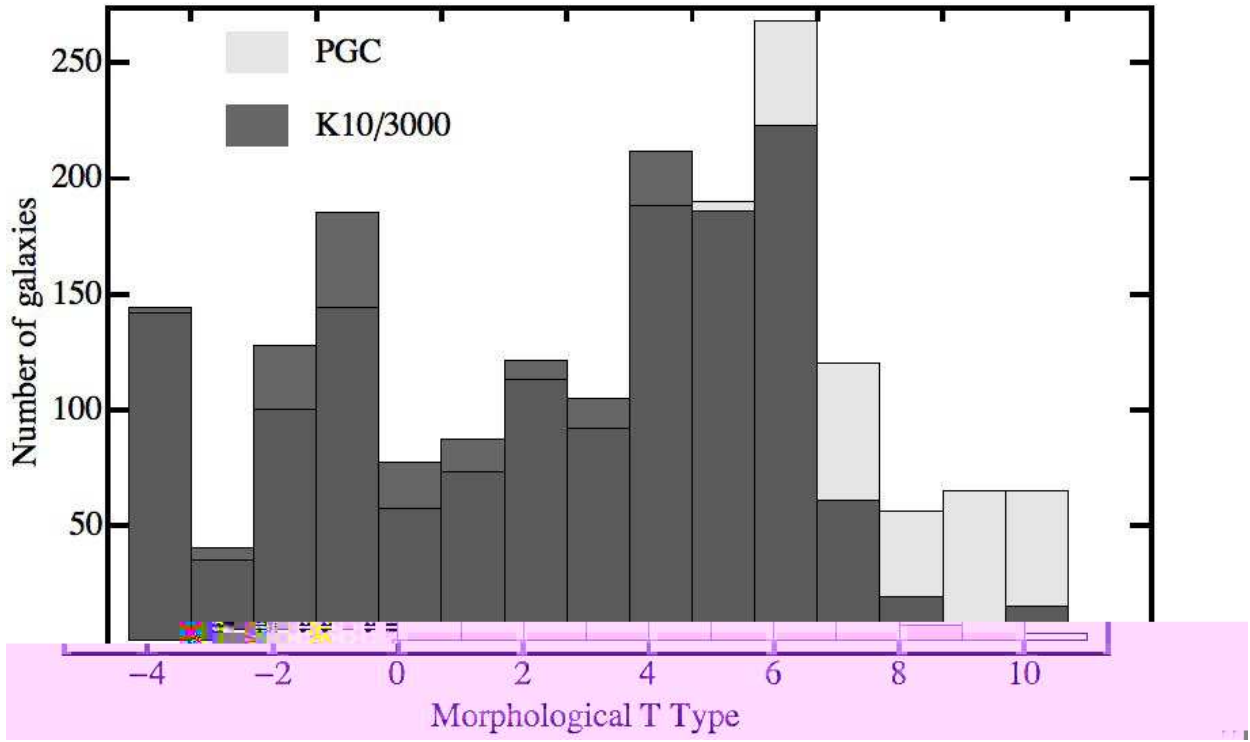}
\caption{ Histograms showing the  distribution of morphological types for a $B$-band selected sample of 1711 nearby  PGC galaxies (V$_{\mathrm gsr}$ ${\le}$ 3000\,km\,s$^{-1}$, ${|b|}$  ${>}$ 10$^{o}$, m$_{B}$ ${\le}$ 13.5 mag) versus the distribution of morphological types for 1610 galaxies in the K10/3000 sample, contained within the same volume.}
\label{default}
\end{center}
\end{figure}

\begin{figure}
\epsscale{1.0}
\begin{center}
\plotone{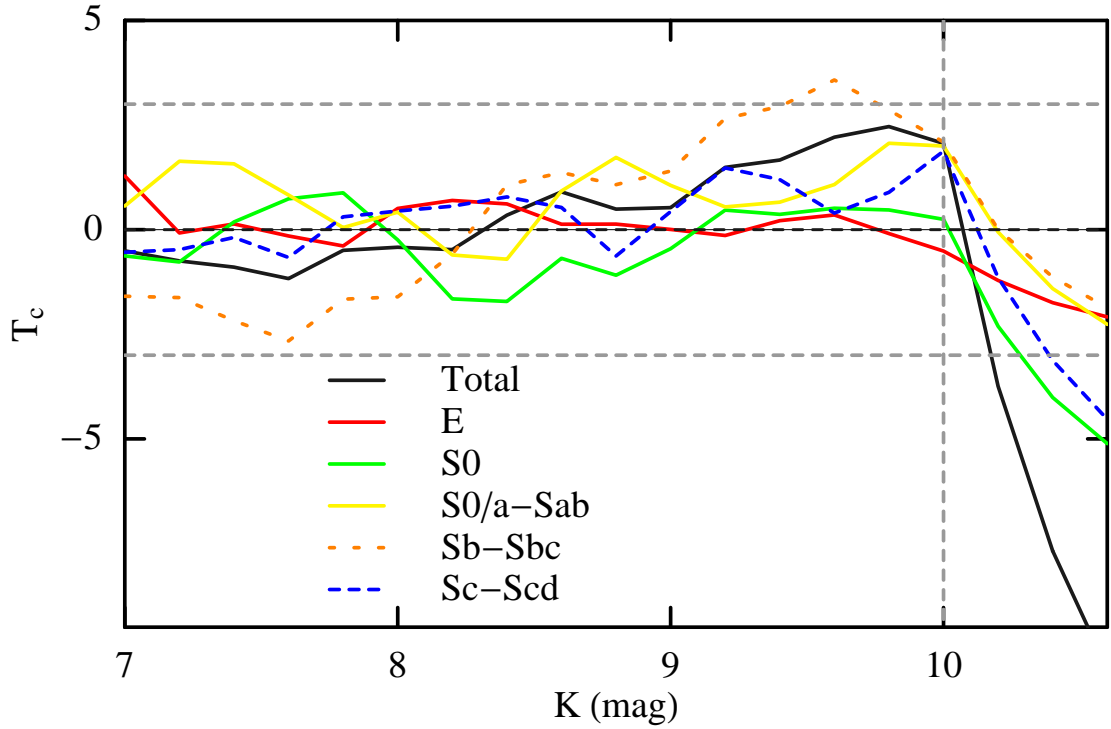}
\caption{Rauzy test for 1601 galaxies in the K10/3000 sample. 
The test is applied to subsamples of the 2MASS XSC
defined at progressively fainter apparent magnitudes. In the present
figure these levels were defined at intervals of 0.1 magnitudes. The horizontal
dashed lines show the levels from where the null hypothesis that the
sample is complete can be rejected at 99.9\% level, under the assumption
of Gaussian statistics. The solid line indicates the expectation value
of $T_c$. The vertical dashed line indicates the adopted limit of
$K$=10.  The colored lines indicate the results for the Rauzy test for
different morphological subsamples  indicated in the figure legend.
The null hypothesis can be rejected with high
confidence beyond $K$ $\sim$ 10.1 where the curve for $T_c$ becomes systematically negative.
Both the number counts and the Rauzy tests suggest that the current sample
is not affected by incompleteness for the morphological types we are
considering.}
\label{default}
\end{center}
\end{figure}

\begin{figure}
\epsscale{1.0}
\begin{center}
\plotone{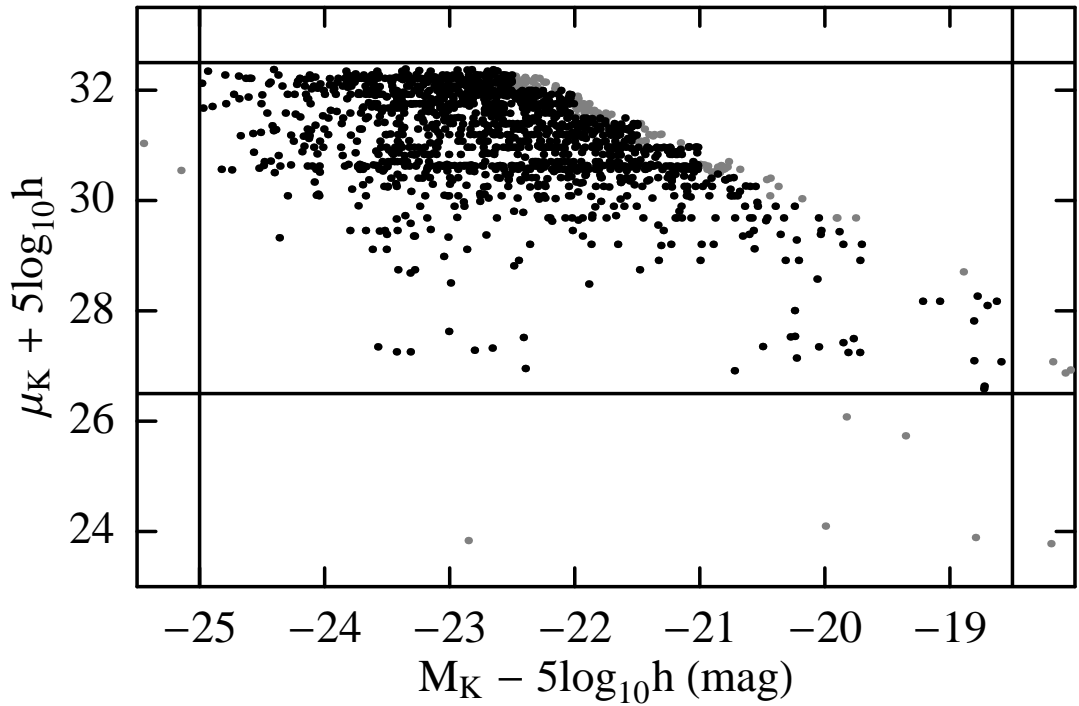}
\caption{ Absolute K magnitude, M$_{K}$ $-$ 5log$_{10}$${\it h}$, versus distance modulus, ${\mu_{K}} $ + 5log$_{10}$${\it h}$, for 1349 galaxies (black dots). Gray dots identify 226 galaxies that are excluded by the binning process (see text for details). The vertical and horizontal lines identify the absolute magnitude and distance modulus limits imposed on the sample prior to computing the total LF (Table 2).}
\label{default}
\end{center}
\end{figure}

\begin{figure}
\epsscale{1.0}
\begin{center}
\plotone{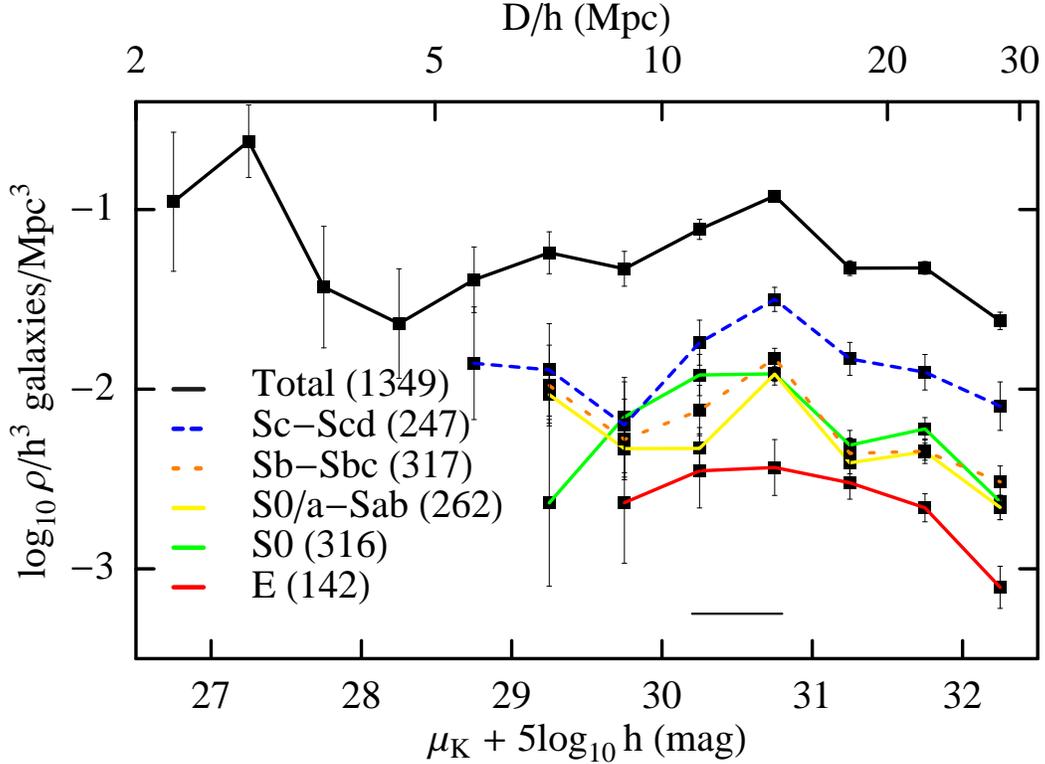}
\caption{ Number density  as a function of distance modulus for 1349  galaxies in the K10/3000 sample segregated by morphological type. For reasons discussed in section 3 all plotted error bars are 50\% larger than calculated using the maximum likelihood method outlined in the Appendix. The horizontal bar under the lowest curve identifies the extent of the Virgo cluster which constitutes only ${\sim}$14\% of the total density in that distance range. Thus, one is cautioned against identifying density enhancements with individual structures because the density is angle averaged over essentially the whole sky. }
\label{default}
\end{center}
\end{figure}

\begin{figure}
\epsscale{0.85}
\begin{center}
\plotone{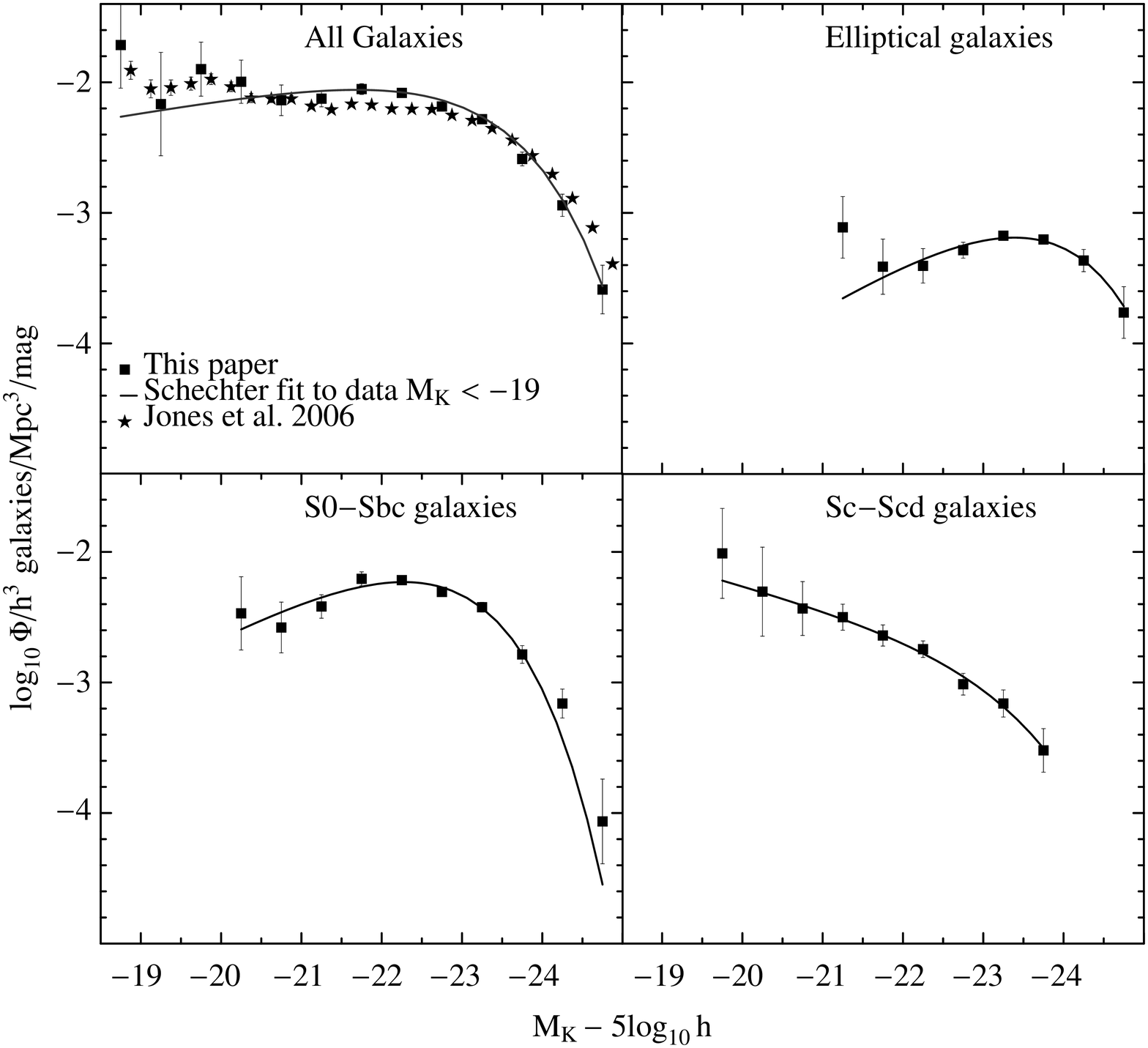}
\caption{${K}$-band isophotal luminosity functions for the K10/3000 sample. 
Filled squares represent the results of this paper and solid lines 
represent Schechter fits to the data (squares)  {\sl Top left panel:} Entire sample of 1349 galaxies.
Increasing incompleteness is expected for M$_{K}$ $-$ 5log$_{10}$${\it h}$ ${\ge}$ $-$19.75 mag
due to the omission of late-type (Sd and later) spiral and dwarf irregular (Im) galaxies. 
Stars identify the $K$-band SWML LF of \cite{Jon06} for a different galaxy sample based on 2MASS isophotal magnitudes and a 
surface brightness dependent correction factor added to emulate total magnitudes. The Schecter fit excludes the first data point at M$_{K}$ $-$ 5log$_{10}$${\it h}$ = $-$18.75. {\sl Top right panel:}   LF for 142 elliptical galaxies. The fit excludes the first data point at M$_{K}$ $-$ 5log$_{10}$${\it h}$ = $-$21.25 which is attributed to the dwarf elliptical sequence. {\sl Lower left panel:}  Combined $K$-band isophotal LF for 904 lenticular (S0) and bulge-dominated spiral (S0/a - Sbc) galaxies.  {\sl Lower right panel:} LF for 247 late-type (Sc--Scd) spiral galaxies. The fit excludes the first data point at M$_{K}$ $-$ 5log$_{10}$${\it h}$ = $-$19.75. For reasons discussed in section 3 all plotted error bars are 50\% larger than calculated using the maximum likelihood method outlined in the Appendix.}
\label{default}
\end{center}
\end{figure}

\begin{figure}
\epsscale{0.85}
\begin{center}
\plotone{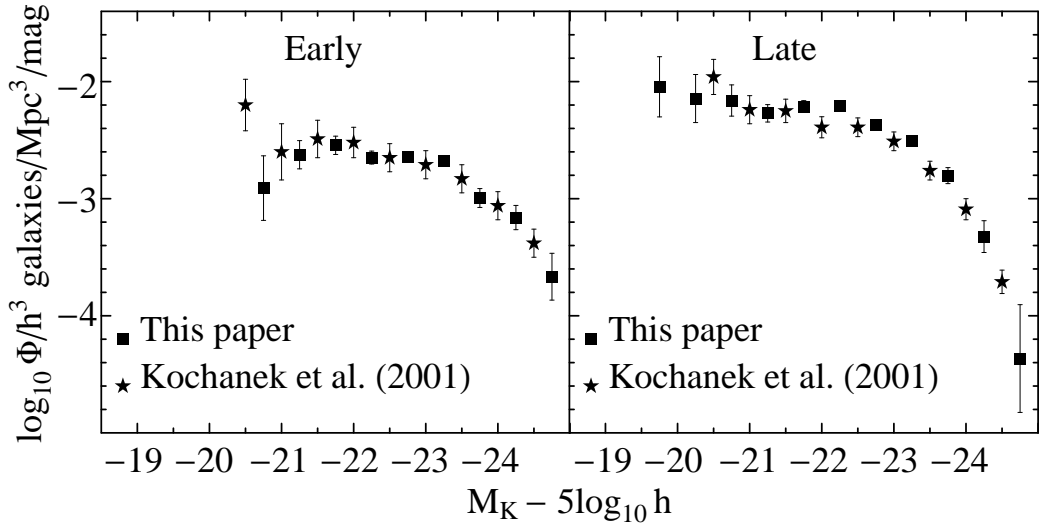}
\caption{ Comparison illustrating the agreement between our results and the cz ${>}$ 2000 km/s, ${K \le}$ 11.25 mag, sample of  \citet{Koc01} when both are divided at T = $-$0.5  such that the early-type subsample includes elliptical and lenticular galaxies and the late-type subsample includes all galaxies classified S0/a and later. The \citet{Koc01} LFs are based on 2MASS isophotal magnitudes. Plotted error bars on the squares are 50\% larger than calculated using the maximum likelihood method outlined in the Appendix.}
\label{default}
\end{center}
\end{figure}

\begin{figure}
\epsscale{1.0}
\begin{center}
\plotone{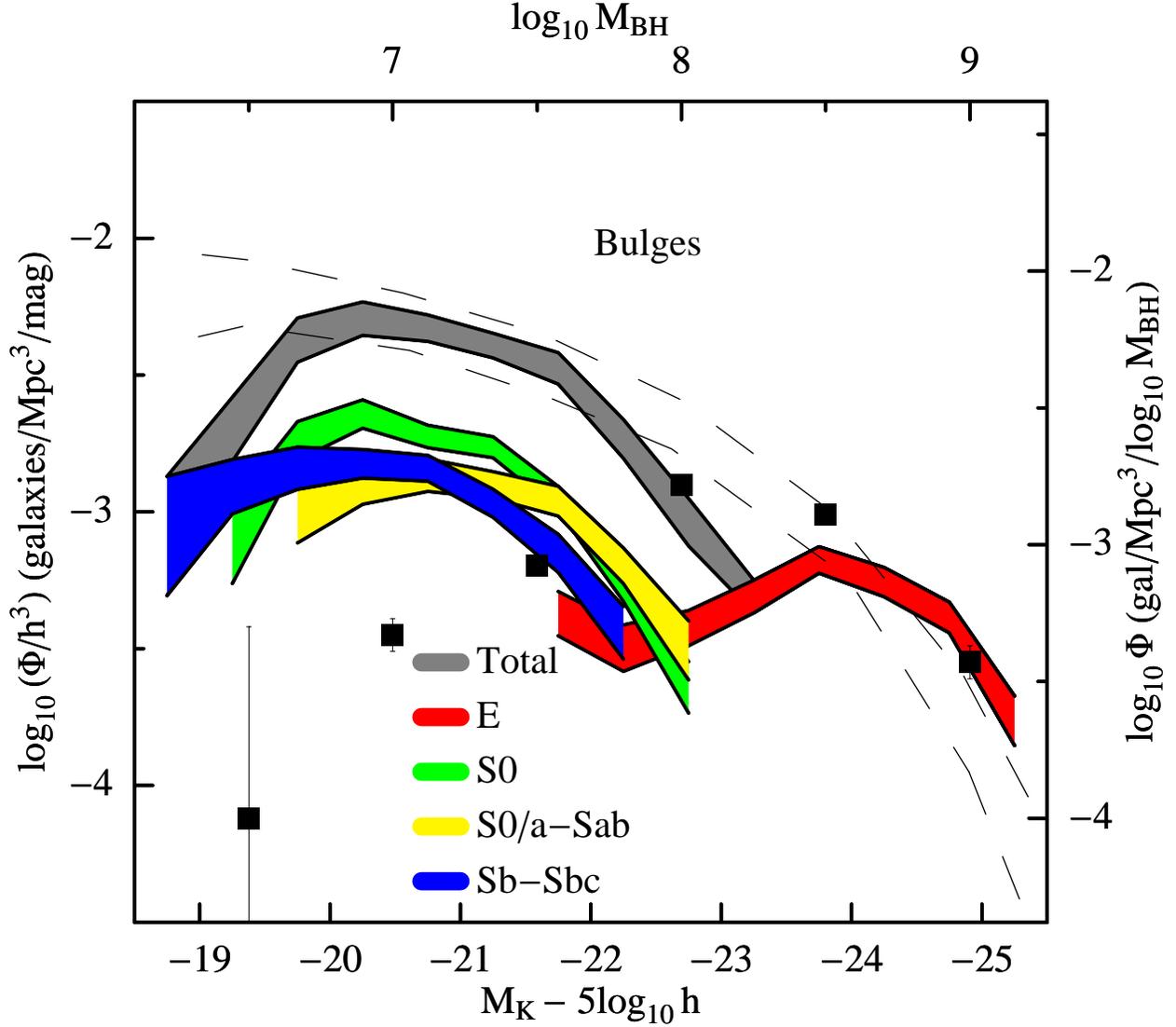}
\caption{Bulge luminosity  functions for galaxies in the K10/3000 sample segregated by morphology. The width of each band reflects the uncertainty
associated with correcting total magnitudes to bulge magnitudes for the S0 - Sbc galaxies and correcting isophotal magnitudes to total magnitudes for the ellipticals.  The total bulge LF merges with the elliptical
LF near M$_{K}$ $-$ 5log$_{10}{\it h}$ ${=}$ $-$23.3 mag.
The upper abscissa and the right hand ordinate show the bulge LFs translated to black hole mass functions using the linear relation of \cite{Mar03}; 
log$_{10}$(M$_{BH}$) = $-$3.757+1.13[(M$_{\odot}$$-$(M$_{K}$ $-$ 5log$_{10}{\it h}$))/2.5] 
and log$_{10}$${\phi(M_{BH})}$ = (2.5${\it h}$$^{3}$/1.13)log$_{10}$${\phi}$(M).
Black hole masses and mass functions are reported in solar units adopting ${\it h}$ = 0.7 and M$_{\odot}$ = 3.32 mag. Dashed lines show a range of BHMFs in the literature as summarized by \cite{Sha09}. Solid squares show the ${\it All~galaxies}$ black hole mass function from \cite{Gra07}.}
\label{default}
\end{center}
\end{figure}

\clearpage








\begin{deluxetable}{lccccccccccc}
\tabletypesize{\scriptsize}
\rotate
\tablecaption{K10/3000 Galaxy Sample \label{tbl-1}}
\tablewidth{0pt}
\tablehead{
\colhead{PGC} & \colhead{NAME} & \colhead{2MASS ID} & \colhead{RA\tablenotemark{a}} & \colhead{DEC} &
\colhead{K\tablenotemark{b}} & \colhead{M$_{K}$\tablenotemark{c}} & \colhead{B\tablenotemark{d}} &
\colhead{LGA\tablenotemark{e}} & \colhead{Morphological\tablenotemark{f}} &
\colhead{Distance\tablenotemark{g}} & \colhead{Quality\tablenotemark{h}}\\
& & &\colhead{Dec. Deg } & \colhead{Dec. Deg.} & \colhead{mag} & \colhead{mag} & \colhead{mag} & & \colhead{T type} & \colhead{Mpc} & \colhead{Distance} \\
\colhead{(1) }& \colhead{(2) }&\colhead{(3) } &\colhead{(4) } & \colhead{(5)} & \colhead{(6)} & \colhead{(7)} & \colhead{(8)} & \colhead{(9) }& \colhead{(10)} & \colhead{(11)} & \colhead{(12)} \\
}
\startdata
 12651	&	NGC1316	&	03224178-3712295	&	50.67	&	-37.21	&	5.69	&	-25.97	&	9.44	&	Z	&	-1.7	&	21.48	&	q	\\
41220	&	NGC4472	&	12294679+0800014	&	187.44	&	8.00	&	5.51	&	-25.66	&	9.27	&	Z	&	-4.8	&	17.14	&	q	\\
30019	&	NGC3147	&	10165363+7324023	&	154.22	&	73.40	&	7.50	&	-25.47	&	11.26	&		&	3.9	&	39.26	&		\\
43296	&	NGC4696	&	12484927-4118399	&	192.21	&	-41.31	&	7.30	&	-25.45	&	11.65	&	Z	&	-3.7	&	35.48	&	q	\\
13505	&	NGC1407	&	03401190-1834493	&	55.05	&	-18.58	&	6.86	&	-25.44	&	10.70	&	Z	&	-4.5	&	28.84	&	q	\\
70090	&	IC1459	&	22571068-3627449	&	344.29	&	-36.46	&	6.93	&	-25.40	&	10.95	&	Z	&	-4.7	&	29.24	&	q	\\
 \enddata
\tablecomments{Table \ref{tbl-1} is published in its entirety in the 
electronic edition of the {\it Astrophysical Journal}.  A portion is 
shown here for guidance regarding its form and content. The galaxies are ordered by absolute K magnitude (column 7).}
\tablenotetext{a}{J2000 coordinates taken from the 2MASS XSC based on the peak pixel \citep{Jar00}.}
\tablenotetext{b}{K magnitude measured within the 20 mag/arc sec${^2}$ elliptical isophote, parameter 
{\tt k${_{-}}$m${_{-}}$k20fe} in the 2MASS XSC. The photometric uncertainty in 
the magnitude is typically 3\%.}
\tablenotetext{c}{Absolute K magnitude computed using the distance in column 11 ($h$ = 0.75). }
\tablenotetext{d}{Total B magnitude taken from the Principal Galaxy Catalog \citep{Pat03}. }
\tablenotetext{e}{Galaxy also listed in the Large Galaxy Atlas  \citep{Jar03} based on the parameter {\tt cc${_{-}}$flg = Z} in the 2MASS XSC.}
\tablenotetext{f}{Morphological T type taken from the Principal Galaxy Catalog \citep{Pat03}.}
\tablenotetext{g}{Distances provided by Tully (2007, private communication). }
\tablenotetext{h}{"q" indicates quality distance (see text for details).}
\end{deluxetable}

\clearpage

\begin{deluxetable}{ccccccc}
\tabletypesize{\scriptsize}
\tablecaption{$K$-band Luminosity Function Binning Parameters\label{tbl-2}}
\tablewidth{0pt}
\tablehead{
\colhead{Sample} & \colhead{M$_{o}$$-$ 5log$_{10}$${\it h}$} & \colhead{M$_{A}$ $-$ 5log$_{10}$${\it h}$} & \colhead{$\mu_{o}$ + 5log$_{10}$${\it h}$} & \colhead{A} & \colhead{B} & \colhead{S} \\
\colhead{} & \colhead{ mag } & \colhead{mag} &  \colhead{mag} & \colhead{mag} & \colhead{ } & \colhead{} \\
\colhead{(1) }& \colhead{(2) }&\colhead{(3) } &\colhead{(4) } & \colhead{(5)} & \colhead{(6)} & \colhead{(7)}  \\
}
\startdata
Total & $-$25 & $-$18.5 & 26.5 &  13 & 12 & 17 \\
Elliptical & $-$25 & $-$21 & 29.5 &  8 & 6 & 11 \\
S0 & $-$24.5 & $-$20.5 & 29 &  8 & 7 & 11 \\
S0/a - Sab & $-$24.5 & $-$21 & 29 &  7 & 7 & 11 \\
Sb--Sbc & $-$24.5 & $-$20.5 & 29 &  8 & 7 & 11 \\
Sc--Scd & $-$24 & $-$19.5 & 28.5 & 9 & 8 & 11 \\
\enddata

\tablecomments{Following the notation of \cite{Cho86}, the apparent magnitude limit, 
m$_{lim}$ = 10 mag, the bin size 
${\Delta}$ = 0.5 mag, the solid angle  ${\Omega}$ = 3.3${\pi}$ steradians and the upper bound on the distance modulus, $\mu_{B}$ + 5log$_{10}$${\it h}$ = 32.5, for all samples}. 

\end{deluxetable}
\clearpage

\begin{deluxetable}{cclccclc}
\tabletypesize{\scriptsize}
\tablecaption{$K$-band Luminosity Functions\label{tbl-3}}
\tablewidth{0pt}
\tablehead{
\colhead{Sample}&\colhead{M$_{K}$ $-$ 5log$_{10}$${\it h}$}&\colhead{ log$_{10}$${\phi}$/${\it h}$$^{3}$ }&\colhead{N}&\colhead{Sample}&\colhead{M$_{K}$ $-$ 5log$_{10}$${\it h}$}&\colhead{ log$_{10}$${\phi}$/${\it h}$$^{3}$ }& \colhead{N} \\
\colhead{}&\colhead{ mag }&\colhead{Gal\,Mpc$^{-3}$ mag$^{-1}$}&\colhead{Gal}&\colhead{}&\colhead{ mag }&\colhead{Gal\,Mpc$^{-3}$ mag$^{-1}$}&\colhead{Gal}\\
\colhead{(1)}&\colhead{ (2) }&\colhead{(3)}&\colhead{(4)}&\colhead{(1)}&\colhead{ (2) }&\colhead{(3)}&\colhead{(4)}\\
}
\startdata
Total&$-$24.75&$-$3.58 ${\pm}$ 0.12&12&Elliptical&$-$24.75& $-$3.76 ${\pm}$ 0.13&8\\
& $-$24.25&$-$2.94 ${\pm}$ 0.06&53&T ${\le}$ $-$3.6&$-$24.25& $-$3.36 ${\pm}$ 0.06$^{*}$&20 \\ 
&$-$23.75&$-$2.59 ${\pm}$ 0.04&120&&$-$23.75&$-$3.20 ${\pm}$ 0.02$^{*}$&29\\ 
&$-$23.25&$-$2.28 ${\pm}$ 0.02&242&&$-$23.25&$-$3.17 ${\pm}$ 0.03$^{*}$&31\\ 
 &$-$22.75 & $-$2.18 ${\pm}$ 0.02 & 303 & &$-$22.75 & $-$3.28 ${\pm}$ 0.04$^{*}$ & 24 \\ 
 &$-$22.25 & $-$2.08 ${\pm}$ 0.02 & 273 & &$-$22.25 & $-$3.40 ${\pm}$ 0.09 & 14 \\ 
 &$-$21.75 & $-$2.05 ${\pm}$ 0.03 & 174 & &$-$21.75 & $-$3.41 ${\pm}$ 0.14 & 8 \\ 
 &$-$21.25 & $-$2.13 ${\pm}$ 0.04 & 97& &$-$21.25 & $-$3.11 ${\pm}$ 0.16 & 8  \\
 &$-$20.75 & $-$2.14 ${\pm}$ 0.08 & 33 \\
 &$-$20.25 & $-$2.00 ${\pm}$ 0.11& 18 \\
 &$-$19.75 & $-$1.90 ${\pm}$ 0.14 & 12 \\
 &$-$19.25 &$-$2.17 ${\pm}$ 0.26 & 3 \\
 &$-$18.75 & $-$1.70 ${\pm}$ 0.22 & 5 \\
 \tableline
 \tablebreak
Lenticular&$-$24.25&$-$3.55 ${\pm}$ 0.11&13&Sa--Sab&$-$24.25&$-$4.06 ${\pm}$ 0.21&4\\
$-$3.5 ${\le}$ T ${\le}$ $-$0.6  &$-$23.75 & $-$3.41 ${\pm}$ 0.09 & 18 & $-$0.5 ${\le}$ T ${\le}$ 2.4 & $-$23.75 & $-$3.26 ${\pm}$ 0.07 &  25 \\
&$-$23.25 & $-$2.84 ${\pm}$ 0.03$^{*}$ & 66 && $-$23.25 & $-$2.97 ${\pm}$ 0.04$^{*}$ & 49  \\
&$-$22.75 & $-$2.75 ${\pm}$ 0.02$^{*}$ & 81 && $-$22.75 & $-$2.80 ${\pm}$ 0.02$^{*}$ & 73  \\
&$-$22.25 & $-$2.73 ${\pm}$ 0.03$^{*}$ & 64 && $-$22.25 & $-$2.76 ${\pm}$ 0.03$^{*}$ & 57 \\
&$-$21.75 & $-$2.60 ${\pm}$ 0.05 & 48 && $-$21.75 & $-$2.69 ${\pm}$ 0.05 & 38 \\
&$-$21.25 & $-$2.76 ${\pm}$ 0.09 & 22 && $-$21.25 & $-$3.00 ${\pm}$ 0.11 & 13 \\
&$-$20.75 & $-$3.10 ${\pm}$ 0.22 & 4 &&  $-$20.75 & $-$3.08 ${\pm}$ 0.26 &  3 \\
 \tableline
Sb--Sbc & $-$24.25 & $-$3.43 ${\pm}$ 0.10 &  17 & Sc--Scd  & $-$23.75 & $-$3.52 ${\pm}$ 0.11 & 14 \\
2.5 ${\le}$ T ${\le}$ 4.4 &$-$23.75 & $-$3.14 ${\pm}$ 0.06 & 33 & 4.5 ${\le}$ T ${\le}$ 6.5 &$-$23.25 & $-$3.16 ${\pm}$ 0.07 & 32 \\
&$-$23.25 & $-$2.90 ${\pm}$ 0.04 & 58 &  &$-$22.75 & $-$3.01 ${\pm}$ 0.06 & 45 \\
 &$-$22.75 & $-$2.79 ${\pm}$ 0.03$^{*}$ & 74 & &$-$22.25 & $-$2.74 ${\pm}$ 0.04$^{*}$ & 55 \\
 &$-$22.25 & $-$2.60  ${\pm}$ 0.02$^{*}$ & 77  & &$-$21.75 & $-$2.64 ${\pm}$ 0.05 & 42\\
 &$-$21.75 & $-$2.75 ${\pm}$ 0.06 & 34& &$-$21.25 & $-$2.50 ${\pm}$ 0.07 & 35 \\
&$-$21.25 & $-$2.90 ${\pm}$ 0.10 & 17 & &$-$20.75 & $-$2.43 ${\pm}$ 0.14 & 12 \\
 &$-$20.75 & $-$2.91 ${\pm}$ 0.20 & 5 &   &$-$20.25 & $-$2.30 ${\pm}$ 0.23 & 5 \\
   & & &   &                       &$-$19.75 & $-$2.01 ${\pm}$ 0.23 & 6 \\
\enddata

\tablecomments{The sum of the number of galaxies within each Hubble type does not add up to the total number of galaxies because of the binning procedure inherent to the Choloniewski method and the fact that every plotted point is defined by a bin containing at least 3 objects. The uncertainties on the LF values are calculated using the maximum likelihood method outlined in the Appendix. Asterisks identify uncertainties that are less than 1/(2.3${\sqrt N}$) suggesting that the galaxy distribution is not Poisson but rather clumped in space and luminosity (section 3; Appendix). Galaxies are segregated 
according to morphological T types; a numerical system developed by \cite{deV59} and \citet{deV91}. The low luminosity end of the total LF, M$_{K}$ $-$ 5log$_{10}$${\it h}$ ${\ge}$ $-$19.75 mag, is defined by very late-types (T ${\ge}$ 6.5) of which there are too few to define a LF separately.}

\end{deluxetable}

\clearpage

\begin{deluxetable}{cccc}
\tabletypesize{\scriptsize}
\tablecaption{$K$-band Luminosity Function Fit Parameters\label{tbl-2}}
\tablewidth{0pt}
\tablehead{
\colhead{Sample} & \colhead{$\phi_{*}$/${\it h}$$^{3}$} & \colhead{M$_{*}$$-$ 5log$_{10}$${\it h}$} & \colhead{$\alpha$}   \\
\colhead{} & \colhead{galaxies\,Mpc$^{-3}$ mag$^{-1}$} &  \colhead{mag} & \colhead{} \\
\colhead{(1)} & \colhead{(2)} &  \colhead{(3)} & \colhead{(4)} \\
}
\startdata
Total &  ( 11.5  ${\pm}$ 3.4 ) ${\times}$ 10$^{-3}$ & $-$23.41 ${\pm}$ 0.46 & $-$0.94 ${\pm}$ 0.10  \\
Elliptical & ( 17.6  ${\pm}$ 0.9 ) ${\times}$ 10$^{-4}$ & $-$23.42 ${\pm}$ 0.17 & $-$0.03 ${\pm}$ 0.16  \\
S0 - Sbc & ( 15.7  ${\pm}$ 1.4 ) ${\times}$ 10$^{-3}$ & $-$22.49 ${\pm}$ 0.20 & $-$0.18 ${\pm}$ 0.16  \\
Sc--Scd & ( 15.9  ${\pm}$ 4.8 ) ${\times}$ 10$^{-4}$ & $-$23.33 ${\pm}$ 0.33 & $-$1.41 ${\pm}$ 0.06  \\
\enddata

\tablecomments{Columns (2) - (4) refer to Schechter function parameters (Figure 7). The uncertainties reflect unweighted fits to the data points.}

\end{deluxetable}

\clearpage

\begin{deluxetable}{cc}
\tabletypesize{\scriptsize}
\tablecaption{$K$-band Luminosity Density by Hubble Type\label{tbl-2}}
\tablewidth{0pt}
\tablehead{
\colhead{Sample} &  \colhead{10$^{7}$${\it h}$L$_{\odot}$\,Mpc$^{-3}$} \\
\colhead{(1)} & \colhead{(2)} \\
}
\startdata
Total &  58  ${\pm}$ 12 \\
Elliptical & 8.1  ${\pm}$ 1.7  \\
S0 - Sbc &  34  ${\pm}$ 7 \\
Sc--Scd &  8.0  ${\pm}$ 1.6   \\
\enddata

\tablecomments{The luminosity densties result from integration of the  Schechter functions, defined in Table~4, between $-$25 ${\le}$ M$_{K}$ ${\le}$ $-$19.}

\end{deluxetable}

\clearpage

\begin{deluxetable}{cccccc}
\tabletypesize{\scriptsize}
\tablecaption{Bulge Luminosity Functions\label{tbl-3}}
\tablewidth{0pt}
\tablehead{
\colhead{Sample}&\colhead{M$_{K}$ $-$ 5log$_{10}$${\it h}$} &\colhead{ log$_{10}$${\phi}$/${\it h}$$^{3}$}&\colhead{Sample}&\colhead{M$_{K}$ $-$ 5log$_{10}$${\it h}$}&\colhead{ log$_{10}$${\phi}$/${\it h}$$^{3}$}\\
\colhead{}&\colhead{mag}&\colhead{Gal\,Mpc$^{-3}$ mag$^{-1}$}&\colhead{}&\colhead{mag}&\colhead{Gal\,Mpc$^{-3}$ mag$^{-1}$}\\
\colhead{(1)}&\colhead{ (2) }&\colhead{(3)}&\colhead{(1)}&\colhead{ (2) }&\colhead{(3)}\\
}
\startdata
Total&$-$25.25& $-$3.76 ${\pm}$ 0.09&Elliptical&$-$25.25& $-$3.76 ${\pm}$ 0.09\\
& $-$24.75&$-$3.38 ${\pm}$ 0.05 &T${\le}$$-$3.6&$-$24.75& $-$3.38 ${\pm}$ 0.06\\ 
& $-$24.25&$-$3.26 ${\pm}$ 0.05&&$-$24.25& $-$3.25 ${\pm}$ 0.05\\ 
& $-$23.75&$-$3.17 ${\pm}$ 0.05&&$-$23.75&$-$3.17 ${\pm}$ 0.05\\
& $-$23.25& $-$3.31 ${\pm}$ 0.06& &$-$23.25 & $-$3.31 ${\pm}$ 0.06\\ 
& $-$22.75 & $-$3.03 ${\pm}$ 0.09& &$-$22.75 & $-$3.42 ${\pm}$ 0.06\\ 
& $-$22.25 & $-$2.73 ${\pm}$ 0.07& &$-$22.25 & $-$3.50 ${\pm}$ 0.08\\ 
& $-$21.75 & $-$2.47 ${\pm}$ 0.06& &$-$21.75 & $-$3.37 ${\pm}$ 0.08\\
& $-$21.25 & $-$2.39 ${\pm}$ 0.04\\
& $-$20.75 & $-$2.33 ${\pm}$ 0.05\\
& $-$20.25 & $-$2.29 ${\pm}$ 0.06\\
& $-$19.75 &$-$2.37 ${\pm}$ 0.08\\
& $-$19.25 & $-$2.69 ${\pm}$ 0.11\\
& $-$18.75 & $-$3.09 ${\pm}$ 0.20\\
 \tableline
 \tablebreak
Lenticular&$-$22.75&$-$3.64 ${\pm}$ 0.09&Sa--Sab& $-$22.75 &$-$3.50  ${\pm}$ 0.10\\
$-$3.5 ${\le}$ T ${\le}$ $-$0.6  &$-$22.25 & $-$3.27 ${\pm}$ 0.05& $-$0.5 ${\le}$ T ${\le}$ 2.4 & $-$22.25 & $-$3.20  ${\pm}$ 0.06 \\
&$-$21.75 & $-$2.95 ${\pm}$ 0.04 &&$-$21.75 & $-$2.96  ${\pm}$ 0.05\\
&$-$21.25 & $-$2.76 ${\pm}$ 0.04 && $-$21.25 & $-$2.90  ${\pm}$ 0.05 \\
&$-$20.75 & $-$2.72 ${\pm}$ 0.04  && $-$20.75 & $-$2.86  ${\pm}$ 0.06\\
&$-$20.25 & $-$2.64 ${\pm}$ 0.05 && $-$20.25 & $-$2.88  ${\pm}$ 0.09\\
&$-$19.75 & $-$2.74 ${\pm}$ 0.07 && $-$19.75 & $-$3.00  ${\pm}$ 0.10\\
&$-$19.25 & $-$3.11 ${\pm}$ 0.14 &&   &  \\

\tableline
Sb - Sbc&$-$22.25 &$-$3.44 ${\pm}$ 0.10& & & \\
2.5 ${\le}$ T ${\le}$ 4.4  & $-$21.75 & $-$3.15 ${\pm}$ 0.07&  &  &  \\
&$-$21.25 & $-$2.97 ${\pm}$ 0.05 & & &  \\
&$-$20.75 & $-$2.84 ${\pm}$ 0.05 & & & \\
&$-$20.25 & $-$2.82 ${\pm}$ 0.05  & &  & \\
&$-$19.75 & $-$2.84 ${\pm}$ 0.08 & &  &\\
&$-$19.25 & $-$2.91 ${\pm}$ 0.10& & &  \\
&$-$18.75 & $-$3.09 ${\pm}$ 0.22&&   &  \\

 \tableline

\enddata

\tablecomments{
The uncertainties reflect the ${ \pm}$ 1${\sigma}$ range in the bulge/total luminosity
ratio propagated through the LFs 55 times using a Monte-Carlo method (section 4.4). }

\end{deluxetable}

\clearpage

\begin{deluxetable}{cc}
\tabletypesize{\scriptsize}
\tablecaption{$K$-band Bulge Luminosity Density by Hubble Type\label{tbl-2}}
\tablewidth{0pt}
\tablehead{
\colhead{Sample} &  \colhead{10$^{7}$${\it h}$L$_{\odot}$\,Mpc$^{-3}$} \\
\colhead{(1)} & \colhead{(2)} \\
}
\startdata
Elliptical & 10.7  ${\pm}$ 1.3  \\
S0 - S0/a &  2.6  ${\pm}$ 0.4 \\
Sa - Sab &  2.7  ${\pm}$ 0.6 \\
Sb - Sbc &  1.8  ${\pm}$ 0.3 \\
Total & 17.8 ${\pm}$ 1.5 \\
\enddata

\tablecomments{The luminosity densties result from integration of the bulge LFs, shown in Figure~9, between $-$25 ${\le}$ M$_{K}$ ${\le}$ $-$19.} 

\end{deluxetable}



\clearpage




\end{document}